\documentclass[twocolumn,showpacs,preprintnumbers,amsmath,amssymb,a4paper,superscriptaddress]{revtex4}
\usepackage{subfigure}
\usepackage{amsmath}
\usepackage{graphicx}
\usepackage{epstopdf}
\usepackage{dcolumn}
\usepackage{bm}
\usepackage{multirow}
\usepackage{float}
\usepackage{booktabs}
\usepackage{tabularx}
\usepackage{array}
\usepackage{makecell}
\usepackage{siunitx}
\usepackage{array,mathtools,amssymb,booktabs}
\newcolumntype{C}{>{$}c<{$}}
\AtBeginDocument
{
\heavyrulewidth=.08em
\lightrulewidth=.05em
\cmidrulewidth=.03em
\belowrulesep=.65ex
\belowbottomsep=0pt
\aboverulesep=.4ex
\abovetopsep=0pt
\cmidrulesep=\doublerulesep
\cmidrulekern=.5em
\defaultaddspace=.5em
}

\begin{document}
\title{Confinement-induced resonance (CIR) in classical vs. quantum scattering under 2D harmonic confinement}
\pacs{}
\author{Saeed Samadi}
\email[]{s.samadi1369@gmail.com}
\author{Bahman Farnudi}
\email[]{farnudi@iasbs.ac.ir}
\author{Shahpoor Saeidian}
\email[]{saeidian@iasbs.ac.ir}
\affiliation{Department of Physics, Institute for Advanced Studies in Basic Sciences
(IASBS),
Gava Zang,
Zanjan 45137-66731,
Iran}

\date{\today}

\begin{abstract}
Using complex analysis, we have investigated classical quasi-one-dimensional atom-atom scattering under 2D harmonic confinement with two different interaction potentials (Yukawa and Lennard-Jones) and found that the Confinement-Induced Resonance (CIR) that occurs in the quantum system seems to have a classical analogy. We observed CIR in our classical results in the sense that a clear minimum appeared in the transmission coefficient for the different interaction potentials. We also investigated the changes in the value and position of this minimum by varying the characteristic parameters of the system including the angular momentum $L_z$ along the longitudinal axis.  In the quantum case, it has already been shown that for the zero range Huang potential, CIR occurs only for $L_z = 0$. Our results indicate that classical CIR can also occur for $L_z \ne 0$ using finite range potentials. We have endeavoured to provide extensive physical arguments to explain the observed results and to support the proposed analogy between the classical and quantum regimes where applicable.
\end{abstract}

\maketitle

\section{Introduction}\label{Intro}

Ultracold gases generally  produced by laser cooling \cite{metcalfad1999} and subsequent evaporative cooling \cite{ketterle1997} at temperatures in nanokelvins provide an excellent opportunity for studying quantum phenomena on mesoscopic and macroscopic scales \cite{s64,s67}. The possibility of controlling the properties of many-body systems using two-body coupling via Feshbach resonances \cite{s99} gives us a new opportunity to study many-body phenomena in physics such as quantum phase transition \cite{s91}, quantum Hall effect \cite{s100}, or BEC-BCS crossover \cite{s92}.

Ultracold gases display interesting aspects of many-body systems when their dynamics is confined to low dimensions. Using optical dipole traps \cite{s15} and atom chips \cite{s18}, mesoscopic structures are manufactured in which the atoms are frozen into occupying a single or a few lowest quantum states of a confining potential in one or more dimensions.  The quantum dynamics of such systems is strongly influenced by the geometry of the confinement. In a confined geometry we may observe the so-called confinement induced resonance (CIR) leading to striking phenomena that do not occur in free space \cite{Olshanii1998,Bergeman2003,Haller,Saeidian08,Saeidian12,Saeidian15,Shadmehri16}.

During the collision process in an ultracold gas, virtual transverse excitations imposed by the confining potential can result in CIR. Here, the atom-atom scattering in the confining transverse potential may be approximated by mapping the 3D s-wave zero-energy scattering of bosons onto an effective 1D longitudinal zero-range pseudopotential. The key notion here is the temporary formation of a molecule resulting from collision under confinement. It has been shown ~\cite{Olshanii1998,Bergeman2003} that CIR occurs if the binding energy of the molecule in the excited state of the trap matches the difference between the energy levels of the confining potential. The confining potential and/or interaction potential may then be altered in a small range around the CIR in order to achieve total reflection (no transmission) resulting in an impenetrable gas \cite{Girardeau1960}.

Now to the realm of quantum and classical mechanics that provide different descriptions of the physical world. In classical physics, particle motion is restricted to classically-allowed regions. Since a particle may not enter a classically-forbidden region, a classical particle cannot travel between disconnected classically-allowed regions. Unlike this classical picture where the trajectory of a particle is deterministic, in quantum mechanics we cannot speak of an actual path that a particle follows when, for example, it tunnels from one classically-allowed region to another. Complex-variable theory has been recently employed by Carl M Bender to provide a familiar, classical picture of this and other well-known quantum effects showing that their qualitative features can be reproduced when the deterministic equations of classical mechanics are extended into the complex plane ~\cite{Proc1}. Many similarities have been found between complex classical mechanics and quantum mechanics ~\cite{QuantEffect,Proc10,Proc11}. In particular, it has been found that a deterministic classical particle whose energy has a small imaginary component can exhibit phenomena that occur in association with quantum mechanics ~\cite{QuantEffect} resembling phenomena that are associated with the uncertainty principle.

Moreover, numerical studies ~\cite{QuantEffect} have shown that a classical particle with complex energy in a double-well potential can exhibit tunneling-like behaviour. Multiple-well potentials have also been studied showing that a classical particle with complex energy in a periodic potential can exhibit a kind of band structure. Other remarkable properties of complex classical trajectories have also been studied ~\cite{Proc5,Proc6,Proc7,Proc9} including the energy discretization of a harmonic oscillator, the complex behaviour of a pendulum, the Euler equations for rigid body rotation ~\cite{Proc10,Proc11}, and the complex extension of chaotic behavior ~\cite{Proc18}.

Inspired by this significant body of work initiated by Bender and extended by many other researchers, we sought to explore the possibility of finding an analogue for the CIR phenomena in real as well as complex classical mechanics. To this end we carried out numerical calculations for the collision of two particles under a two-dimensional harmonic confining potential. We solved Hamilton's equations for this system for two different inter-particle interaction potentials, the Lennard-Jones and Yukawa potentials, and observed resonance phenomena that appear to be similar to the quantum CIR case where under certain conditions the transmission coefficient shows a minimum. Changes in the value and position of this minimum were studied by varying different parameters of the system that included the energy of the incoming particle, potential range, and the angular momentum $L_z$ along the longitudinal axis. In the quantum case, it has already been shown that for the zero range Huang potential, CIR occurs only for $L_z = 0$ \cite{Moore2003}. Our results indicate that classical CIR can also occur for $L_z \ne 0$ using finite range potentials.

Unlike the quantum case where a distinction is made between distinguishable and indistinguishable particles, our classical results clearly only apply to distinguishable particles. We have provided physical arguments to interpret our observed results and to support the analogy between the classical and quantum regimes wherever quantum results were available. A physical explanation has also been put forward to show that the observed classical CIR is similar to the quantum CIR in origin and type.

This paper is organized as follows. In Section \ref{PTsymmetry} we review the pre-requisite complex mechanics and the concept of $\mathcal{PT}$-symmetry. In Section \ref{HamiltonianAndScattering} we write down the Hamiltonian equations for the two-body scattering problem in a two-dimensional confining potential.  Section \ref{Numerical} describes the numerical approach we have used in the classical (\ref{Classical}) and quantum (\ref{Quantum}) cases. In Section \ref{Results} we present our results, in section \ref{Discussion} we provide extensive physical arguments to explain them, and in the final section we summarize our results and draw some conclusions.

\section{$\mathcal{PT}$-symmetry and the complex Hamiltonian}\label{PTsymmetry}

First proposed by Bender and Boettcher in 1998 \cite{thesis36} $\mathcal{PT}$-symmetry in quantum theory is a physical geometric condition, weaker than the mathematical condition of Hermiticity for the Hamiltonian, which nevertheless guarantees the eigenvalue spectrum of the system to be real. The parity operator $\mathcal{P}$ and the time reversal operator $\mathcal{T}$ are defined by their actions on the position and momentum operators as

\begin{align}
\label{PT1}
\mathcal{P}&: \:x\rightarrow -x, \quad p \rightarrow -p, \\
\label{PT2}
\mathcal{T}&: \:x\rightarrow x, \quad p \rightarrow -p,
\quad i \rightarrow -i . \nonumber
\end{align}

The eigenvalue $E$ of $H$ is real if (1) $[ \mathcal{PT} ,H ] =0$ and (2) the $\mathcal{PT}$-symmetry of $H$ is unbroken; that is, if the corresponding eigenfunction of $H$ exhibits $\mathcal{PT}$-symmetry. As an example, for $H=p^2-(ix)^N$ with $N\ge 2$, these two conditions are satisfied and the eigenenergies are real \cite{thesis36}.
This discovery, which was primarily deemed to be mathematical, has now been explored experimentally in several fields of physics such as NMR, solid-state physics, and optics.

$\mathcal{PT}$-symmetry has also been studied in complex classical mechanics \cite{kaushal2001construction}.
By extending the variables $x$ and $p$ of classical mechanics into the complex domain as

\begin{align}
x &=  x_{1} + i x_{2}, \\
p &=  p_{1} + i p_{2},\nonumber
\end{align}
the $\mathcal{P}$ and $\mathcal{T}$ transformations can be written as

\begin{align}
\mathcal{P}&:  x_{1} \rightarrow - x_{1}, \;  x_{2} \rightarrow - x_{2}, \;   p_{1} \rightarrow - p_{1}, \;  p_{2} \rightarrow - p_{2},\qquad \\
\mathcal{T}&:  x_{1} \rightarrow  x_{1}, \;  x_{2} \rightarrow - x_{2}, \;   p_{1} \rightarrow - p_{1}, \;  p_{2} \rightarrow  p_{2},
\; i \rightarrow -i. \nonumber
\end{align}

Recalling Hamilton's equations for the complex version of the Hamiltonian $H = H_{1}(x_{1},p_{1};x_{2},p_{2}) + i H_{2}(x_{1},p_{1};x_{2},p_{2})$ for the case where $H$ is an analytic function ($i.e.$, it satisfies the Cauchy-Riemann condition), we obtain

\begin{align}
 \label{ModHamil1}
 \dot{x}_{1} &= 2 \dfrac{\partial H_{1}}{\partial p_{1}},\quad \dot{p}_{1} = -2 \dfrac{\partial H_{1}}{\partial x_{1}},\\
 \label{ModHamil2}
 \dot{x}_{2} &= -2 \dfrac{\partial H_{1}}{\partial p_{2}},\quad \dot{p}_{2} = 2 \dfrac{\partial H_{1}}{\partial x_{2}} .
 \end{align}

In the case of a $\mathcal{PT}$-symmetric Hamiltonian, the solutions of the above equations for real energies will be $\mathcal{PT}$-symmetric.

We recall that quantum and classical mechanics provide deeply different views of the physical world. While quantum mechanics is a non-local and probabilistic theory in which one cannot speak of an actual and deterministic path for the particle, in classical mechanics the trajectory of the particle is localized and deterministic, being the solution of a local initial value problem for Hamilton's differential equation. Quantum effects such as discreteness of energy and tunnelling are consequences of the non-local nature of the theory and do not have any analogues in classical mechanics where the energy $E$ is a continuous quantity and the trajectories of the particles in motion under the potential $V(x)$ are restricted to classically-allowed regions, where $E\ge V(x)$.

\begin{figure}

\centering
\includegraphics[width=0.48\textwidth]{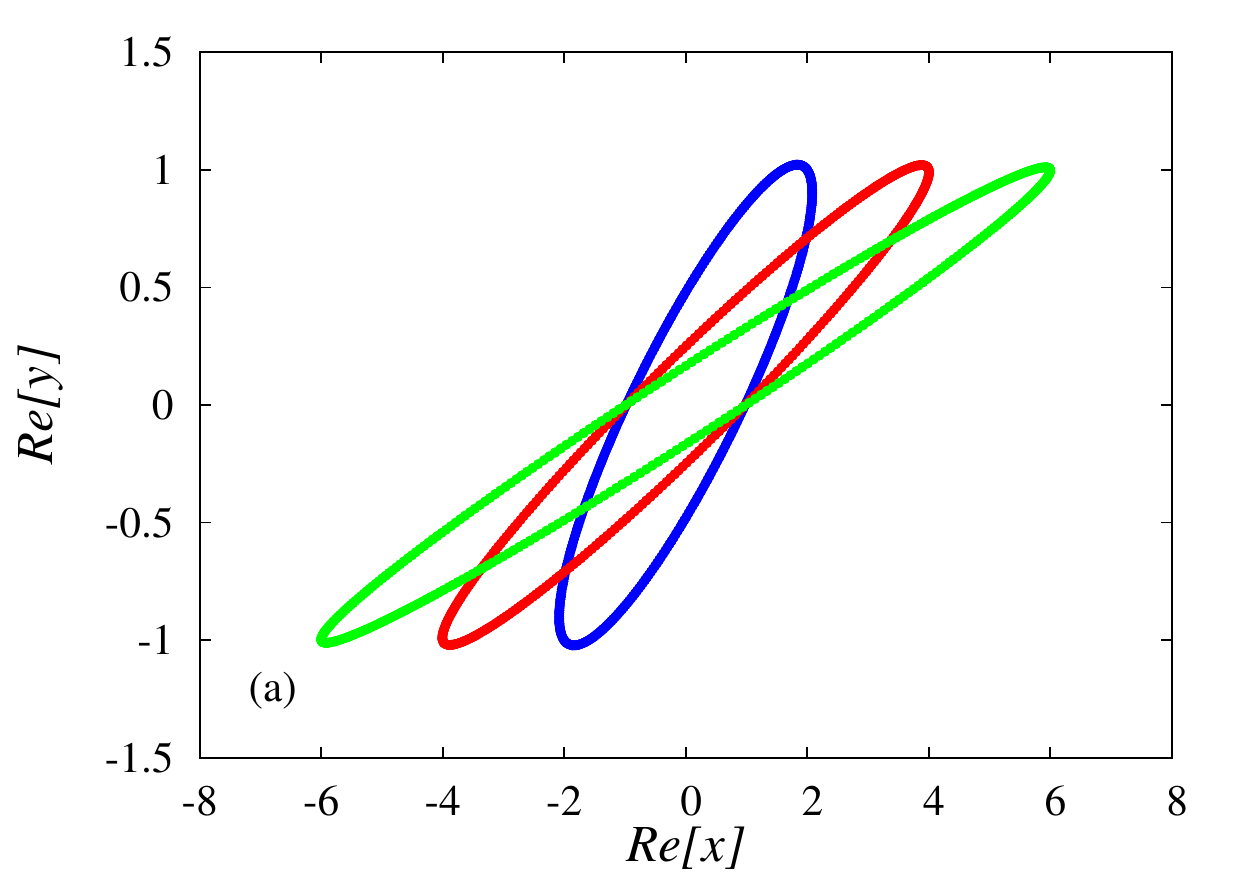}
\includegraphics[width=0.48\textwidth]{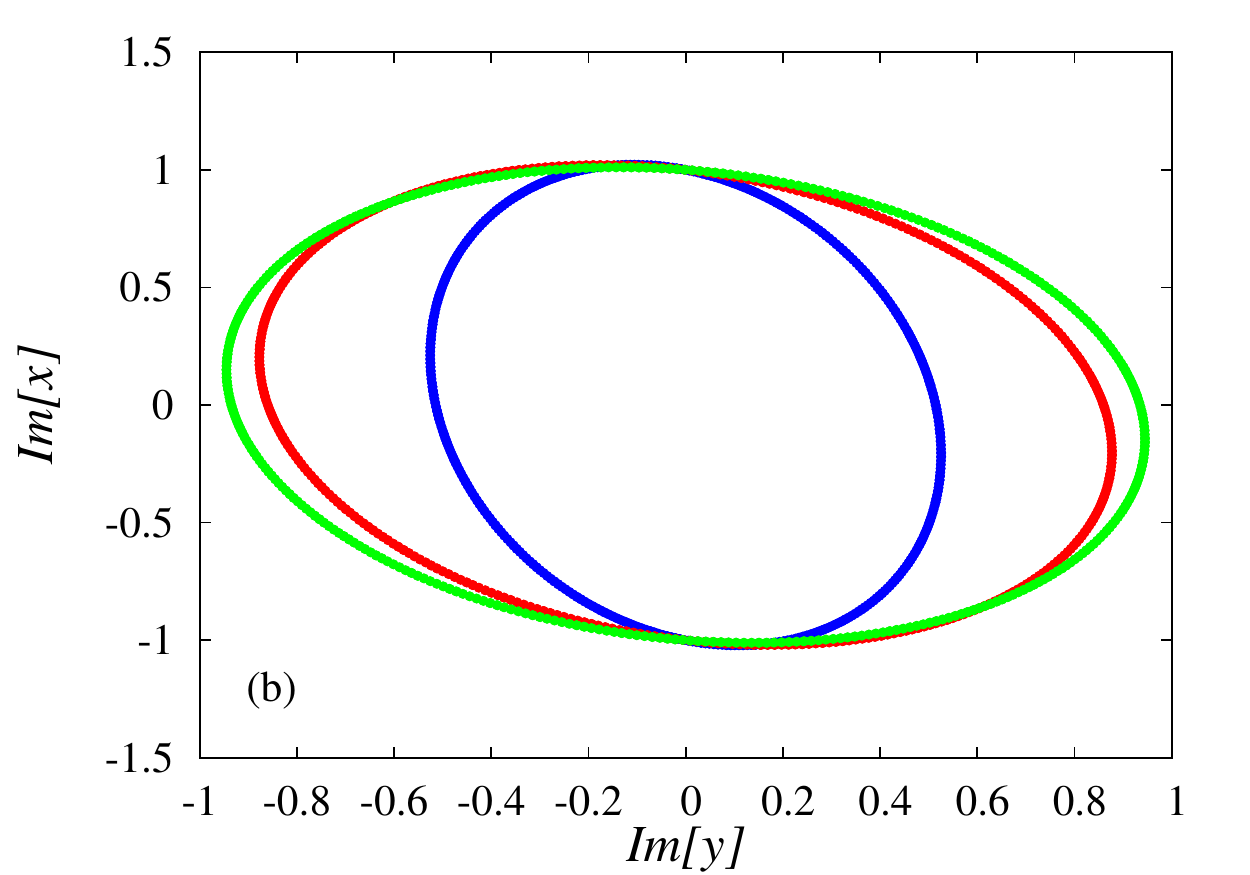}
\includegraphics[width=0.48\textwidth]{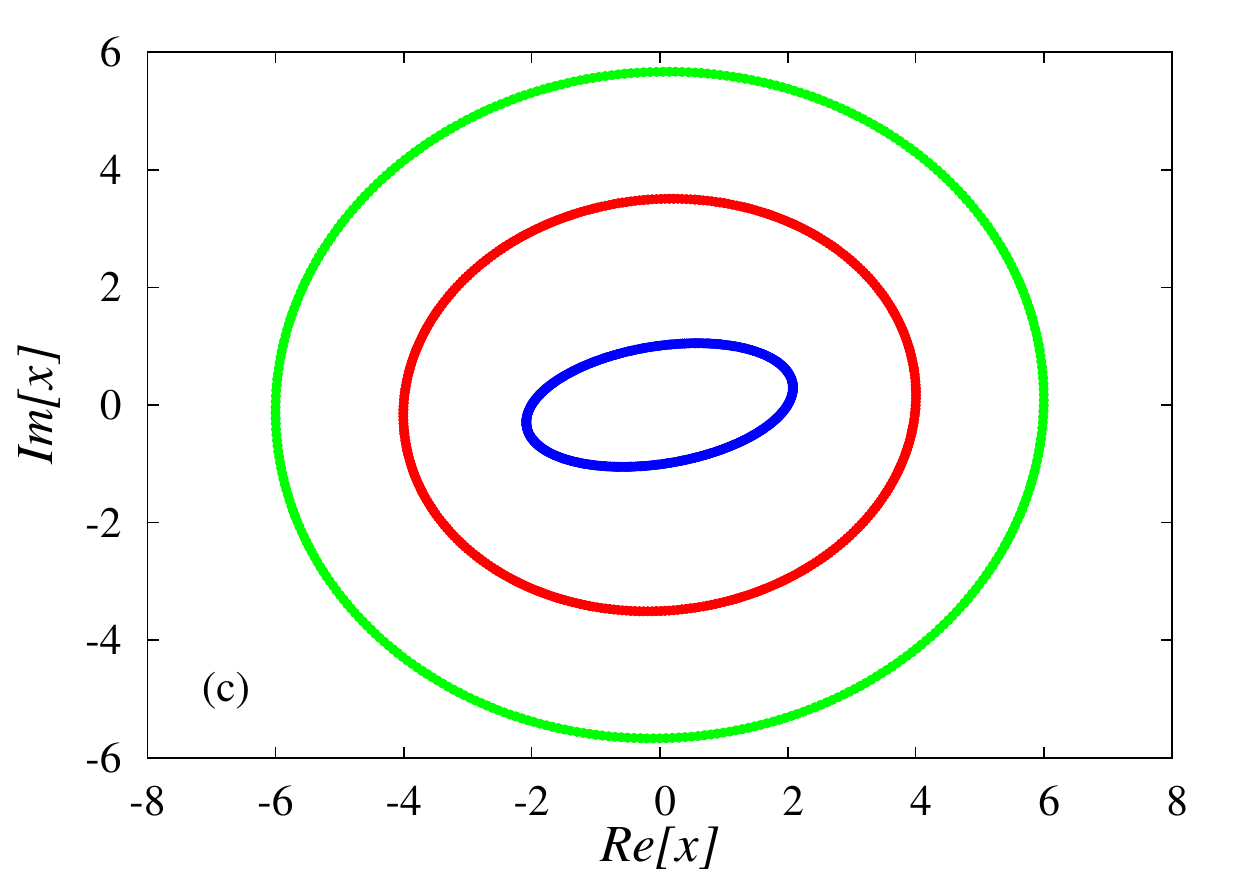}
\includegraphics[width=0.48\textwidth]{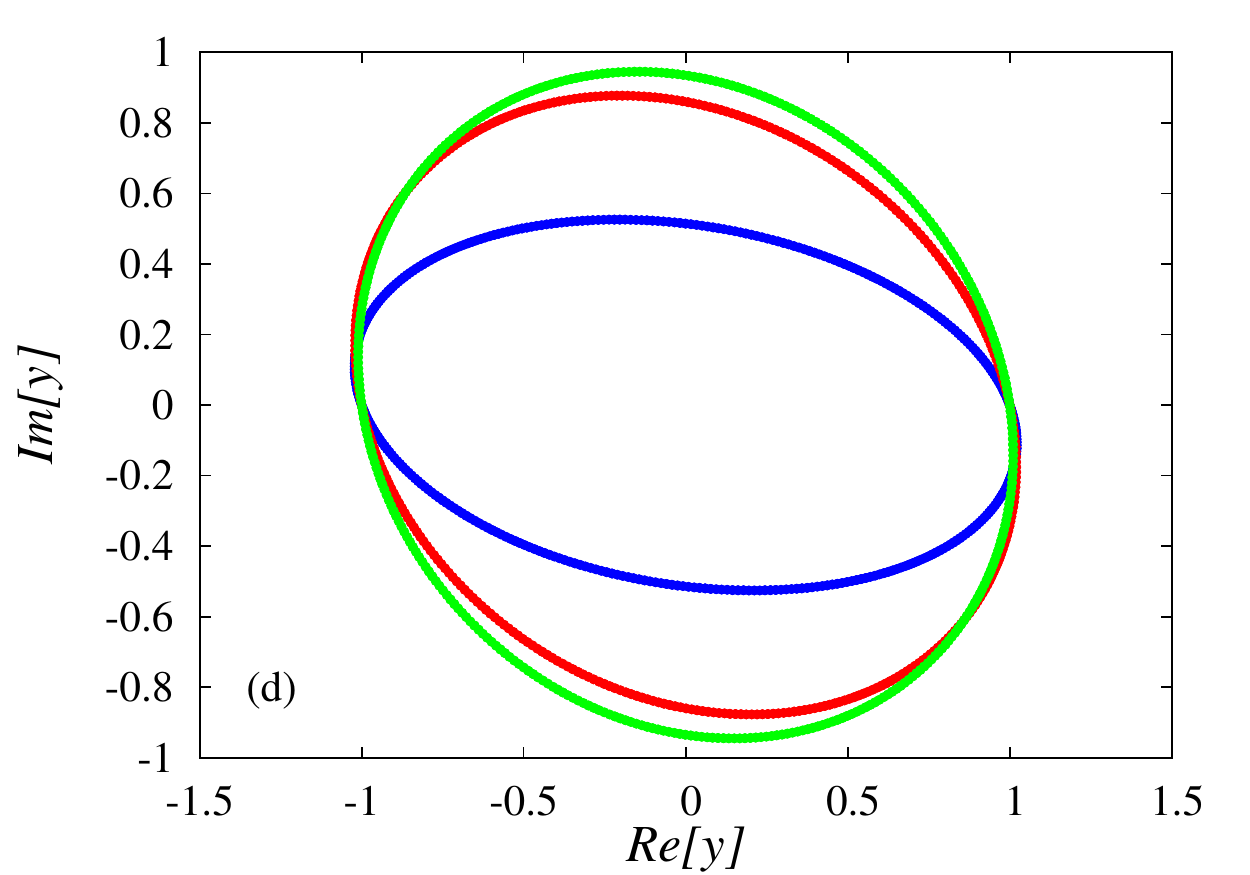}
\caption{(Color online) Different cuts through the 6-dimensional path of a 2D harmonic oscillator in the complex hyperspace for $L_z=1$.} \label{2D orbit}
\end{figure}

However, in some special cases it is possible to make a connection between classical and quantum mechanics. One famous example is the Bohr-Sommerfeld quantization formula (assuming $\hbar = 1$)

\begin{align}
 \oint_c p \, dx \sim (n+\frac{1}{2}) 2\pi,
\end{align}
which gives a semiclassical approximation to the energies of the Hamiltonian, provided that the classical orbit is closed. This condition is satisfied for a harmonic oscillator, even in the complex domain (see Fig.\ref{2D orbit} for the 2D oscillator). The Hamiltonian for a 2D harmonic oscillator (which we use in our problem as a confining potential) is given by

\begin{align}\label{HO2D}
H = \dfrac{{p_{x}}^2}{2 m} + \dfrac{{p_{y}}^2}{2 m} + \dfrac{1}{2} m \omega^2 x^2 + \dfrac{1}{2} m \omega^2 y^2.
\end{align}
Assuming $m=1$ and writing the Hamiltonian in the cylindrical coordinates, we have

\begin{align}\label{2DHarmonicH}
H = \dfrac{\dot{\rho}^{2}}{2} + \dfrac{\rho^{2}\dot{\phi}^2}{2} + \dfrac{\rho^{2}\omega^2}{2}.
\end{align}

By introducing the angular momentum along the z-axis as $L_z=\rho^2\dot{\phi}$, expression (\ref{2DHarmonicH}) can then be written as

\begin{align}
H = \dfrac{\dot{\rho}^{2}}{2} + \dfrac{L_z^{2}}{2\rho^2} + \dfrac{\rho^{2}\omega^2}{2}.
\end{align}

\noindent Accordingly, the turning points for an arbitrary energy are evaluated as

\begin{align}
 \rho_1 = \sqrt{E - \sqrt{E^2 - L_{z}^2}} 
\end{align}

\noindent and

\begin{align}
 \rho_2 = \sqrt{E + \sqrt{E^2 - L_{z}^2}}.
\end{align}

\begin{figure}
\includegraphics[width=0.48\textwidth]{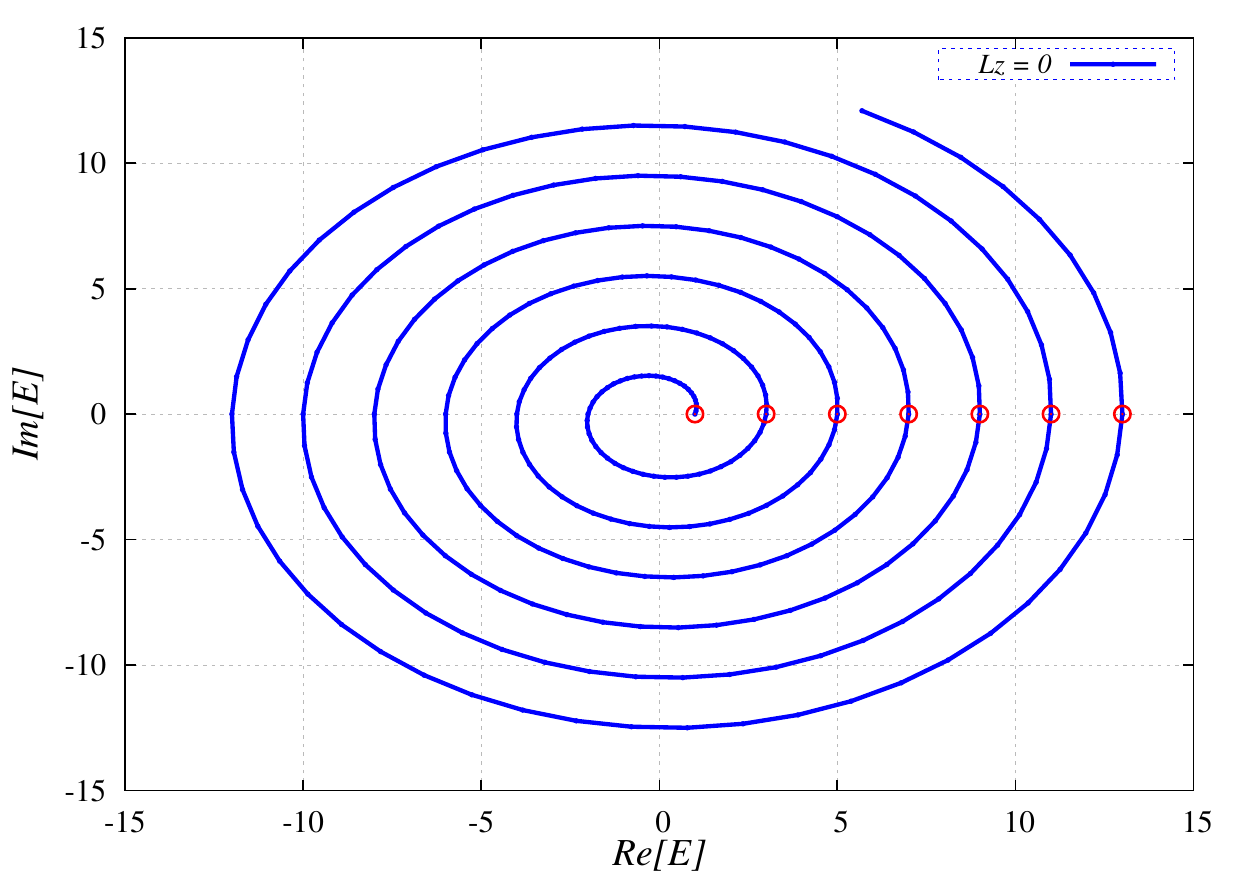}
\caption{\footnotesize (Color online) Representing the energies of a 2D harmonic oscillator on the complex plane for $L_{z}=0$. Open circles show the energies predicted by quantum mechanics.} \label{2D_Riemann}
  \end{figure}

Applying the Bohr-Sommerfeld quantization rule to the 2D harmonic oscillator for a particular $L_z$, we can obtain the energy spectra whose real and positive values correspond to the analogous quantum eigenenergies $E_{\perp} = (2n + |L_z| + 1) \omega$. Figure \ref{2D_Riemann} shows the results for $L_z=0$. The open circles correspond to the quantum energy eigenvalues obtained from the Hamiltonian (\ref{2DHarmonicH}). In the case of quantum scattering in a harmonic waveguide, these discrete energy values are in fact the thresholds of the so-called scattering channels. Since we intended to try and emulate the corresponding classical system and compare our classical results with the quantum ones, we used the above discrete energies as the transverse energy ($E_{\perp}$) of the classical case.

\section{The Hamiltonian and the two-body scattering problem under 2D harmonic confinement}\label{HamiltonianAndScattering}

Let us consider the collision of two particles in three dimensions under a transverse 2D harmonic confining potential characterized by a single frequency $\omega$ for each particle. The corresponding Hamiltonian is given by

\begin{multline}
H = \dfrac{\mathbf{p}_{1}^2}{2m_{1}}  + \dfrac{\mathbf{p}_{2}^2}{2m_{2}} + \dfrac{1}{2} m_{1} \omega^2 (\mathbf{x}_{1}^2 + \mathbf{y}_{1}^2)\\ + \dfrac{1}{2} m_{2} \omega^2 (\mathbf{x}_{2}^2 + \mathbf{y}_{2}^2) + V_{int} (\mathbf{r}_{1} - \mathbf{r}_{2}),
\end{multline}

\noindent where $m_{i}$ is the mass of the $i^{th}$ particle and $V_{int} (\mathbf{r}_{1} - \mathbf{r}_{2})$ is the two-body interaction potential. The Hamiltonian can be separated with respect to the relative ($rel$) and center of mass (CM) coordinates ($\mathbf{r}_{rel}=\mathbf{r}_{1}-\mathbf{r}_{2}, \mathbf{R}_{CM}=m_1\mathbf{r}_{1}+m_2\mathbf{r}_{2}/m_1+m_2$)

\begin{equation}
H = H_{CM} + H_{rel},
\end{equation}

\noindent where

\begin{equation}
H_{CM} = \dfrac{\mathbf{p}_{CM}^2}{2M}  +  \dfrac{1}{2} M \omega^2 \mathbf{x}_{CM}^2 + \dfrac{1}{2} M \omega^2 \mathbf{y}_{CM}^2,
\end{equation}

\begin{equation}\label{Hrel}
H_{rel} = \dfrac{\mathbf{p}_{rel}^2}{2\mu}  +  \dfrac{1}{2} \mu \omega^2 \mathbf{x}_{rel}^2 + \dfrac{1}{2} \mu \omega^2 \mathbf{y}_{rel}^2 + V_{int} (\mathbf{r}_{rel}),
\end{equation}

\noindent and $M=m_{1}+m_{2}$ and $\mu=\frac{m_{1}m_{2}}{m_{1}+m_{2}}$ are the total  and reduced masses, respectively.

Since the solutions for $H_{CM}$, which describes a 2D harmonic oscillator are known, from this point on, we focus on $H_{rel}$ which describes the scattering of a single effective particle with reduced mass $\mu$ and energy $E$, from a short range scatterer $V_{int}(\mathbf{r})$ at the origin, under 2D transverse harmonic confinement with frequency $\omega$. 
Far from the origin, since $V_{int}(\bold{r})$ is short range it can be neglected. Therefore, far from the origin, the Hamiltonian $H$ may be written as $H=H_{\perp}+H_{||}$, where
\begin{equation}\label{Hrel}
H_{\perp} = \dfrac{p_{x}^2}{2\mu}  +  \dfrac{p_{y}^2}{2\mu}+\dfrac{1}{2} \mu \omega^2 (x^2 +y^2)
\end{equation}
and
\begin{equation}\label{Hrel}
H_{||} = \dfrac{p_{z}^2}{2\mu}.
\end{equation} 
$H_{\perp}$ and $H_{||}$ describe simple harmonic oscillation on the $x-y$ plane and free motion along the $z-$ axis, respectively.
The corresponding energy can then be written  as the sum of the transverse energy $E_{\perp}$ (of the harmonic oscillation) and longitudinal energy $E_{||}$ (of the free motion along the $z-$axis), $E=E_{\perp}+E_{||}$. We are now going to treat this problem in the classical and quantum cases.
\subsection{Scattering in the classical case}\label{ScatteringClassical}

For the transverse energy $E_{\perp}$ we choose the values obtained from the Bohr-Sommerfeld quantization rule ($E_{\perp}=E-E_{||}=\hbar \omega (2n+|L_z|+1)>0$) in order to simulate the concept of channels in quantum mechanics.

We carried out a large number of numerical calculations simulating the collision process by solving the modified Hamilton's equations using the \textit{Runge-Kutta-Fehlberg} method ~\cite{Hoffman}. Figure \ref{Trajectories} shows typical resulting trajectories for the Yukawa interaction potential. There were two cases, the particle being reflected (a) or transmitted (b).

\begin{figure}
\includegraphics[width=0.48\textwidth]{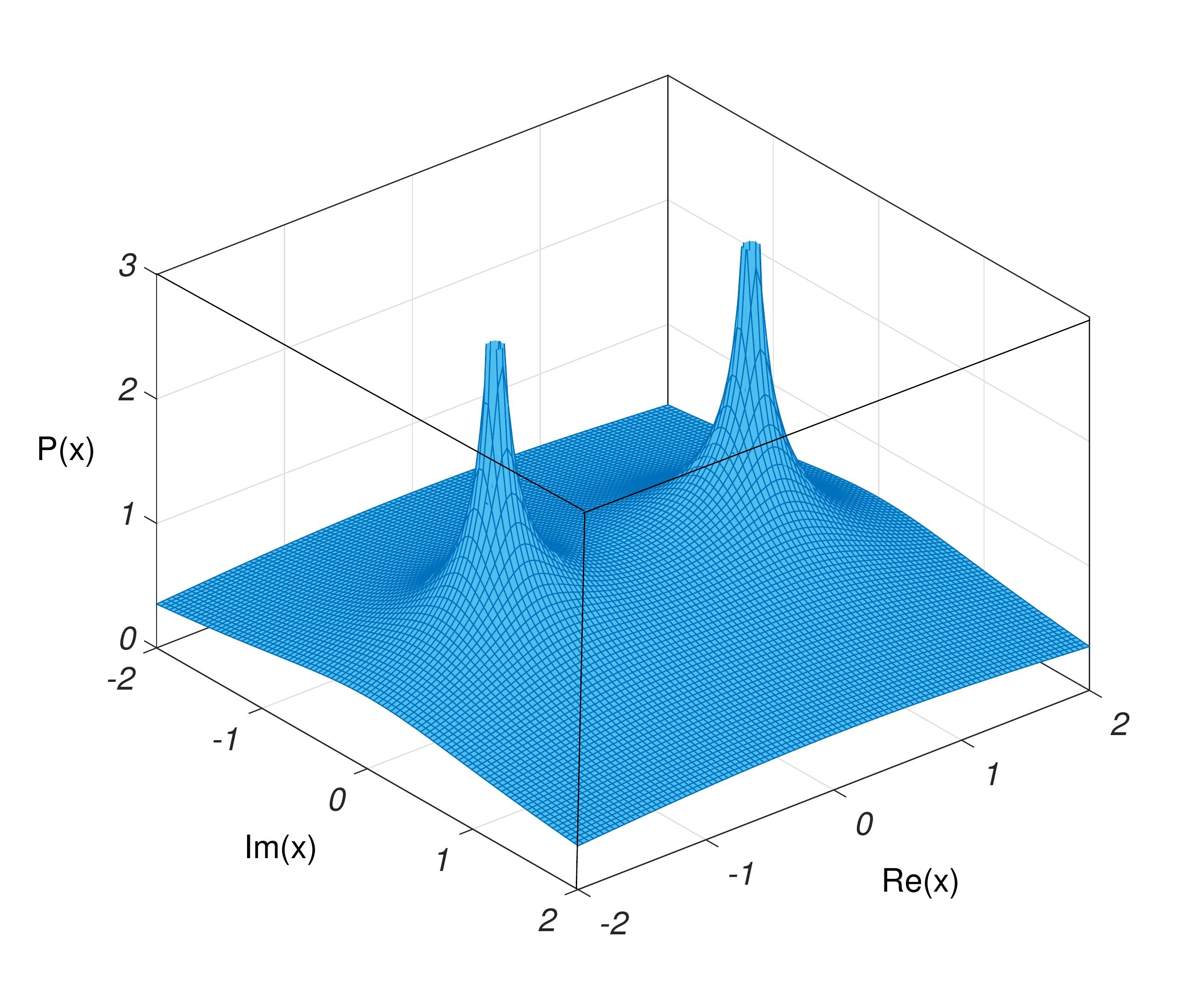}
\caption{\footnotesize (Color online) The relative spatial distribution $P(\textbf{r})=P(x)$ of the incident particle's initial position $\textbf{r}=(x, y, z)$ on the complex plane of $x$ for $L_z=0$,   $y=0$ and $z=-\infty$.} \label{RelProCom}
  \end{figure}

\begin{figure}[h!]
\centering
\subfigure{
\includegraphics[scale=0.7]{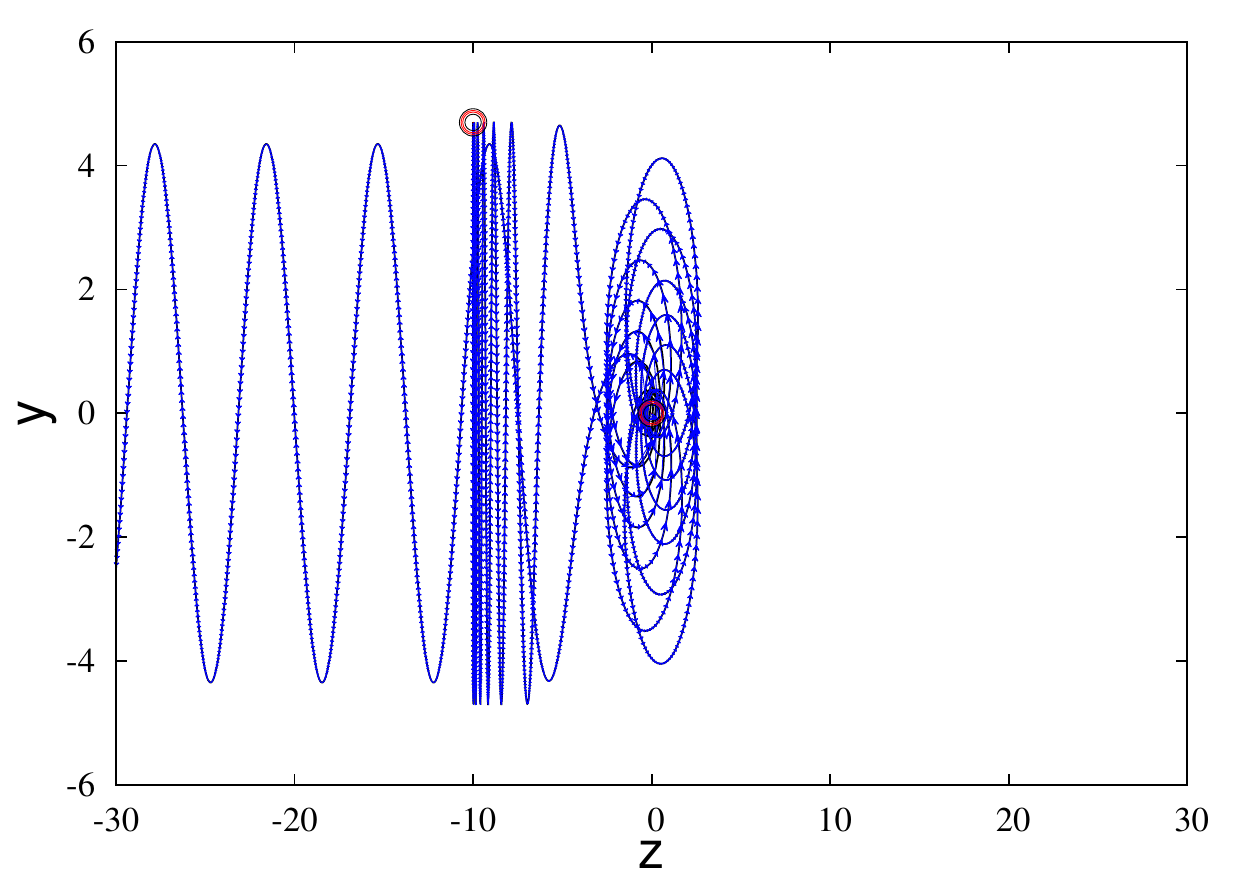}}
\subfigure{
\includegraphics[scale=0.7]{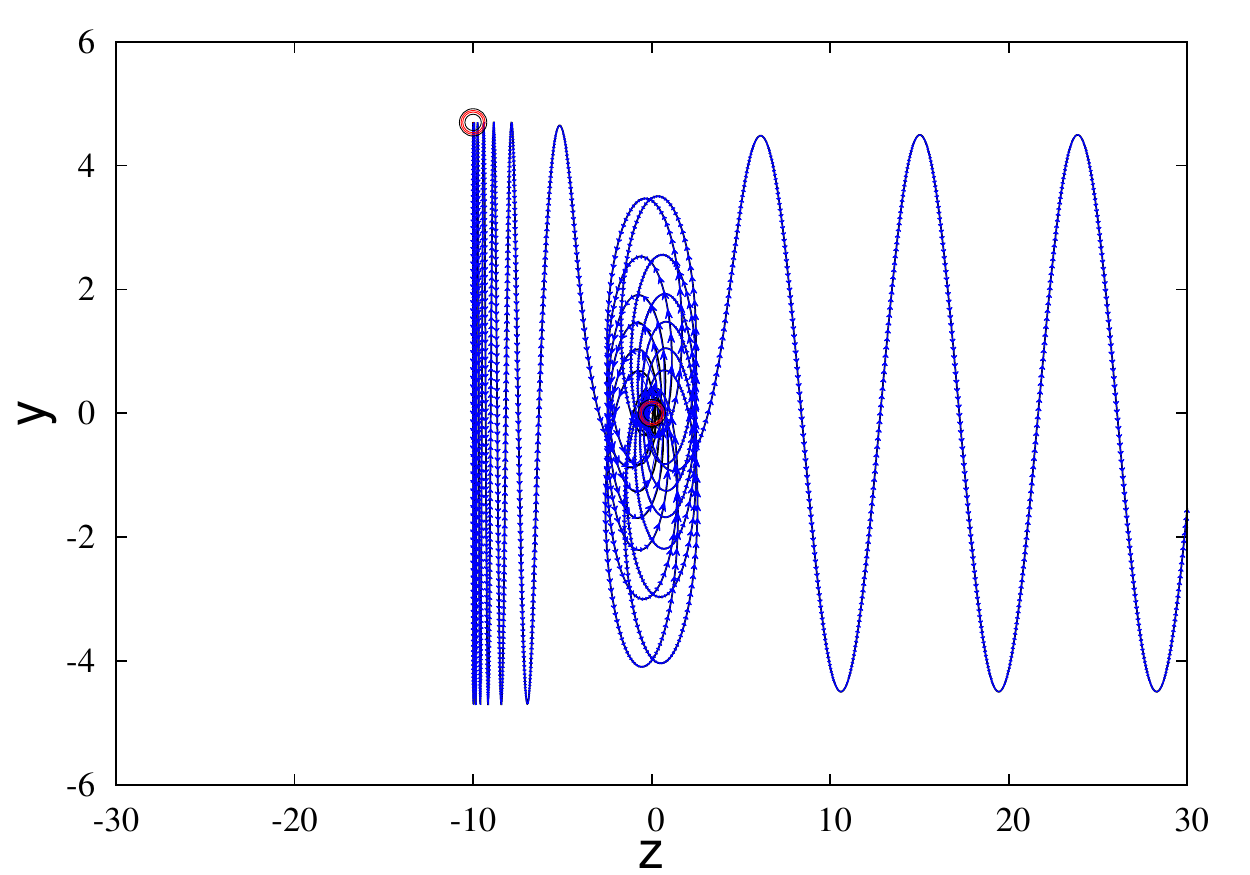}}
\caption{\label{Trajectories} (Color online) The typical trajectory of a particle confined by a 2D harmonic oscillator and interacting with the Yukawa interaction potential (\ref{yuk}), for $V_0=50$, $r_0=1$, $\omega=1$, $E_{\perp}=11$,  $E_{||}=1.0\times 10^{-6}$, and $L_z=0$. The particle is originally at $z = -10$ and is reflected to $-\infty$ (a) or transmitted to $+\infty$ (b).}  
\end{figure}

We used a total of $N=1600$ different initial conditions for particular values of the energies $E_{\perp}$ and $E_{||}$, angular momentum $L_z$, and interaction potential depth $V_0$. The initial position of the incident particle was set far from the scatterer ($z<<0$).  We chose all the allowed orbits (see Fig. \ref{2D orbit}) of the confining potential randomly and with equal weight. Therefore, the spatial distribution of the initial position $\textbf{r} = (x,y,z)$ of the incident particle will be given by $P(\textbf{r})  = \frac{1}{v(\textbf{r})}$, $v(\textbf{r})$ being the speed of the particle \cite{Bender 2010}.   Because of the axial symmetry of the system, $P(\textbf{r})$ is symmetric about the $z-$axis.   In Fig. \ref{RelProCom} we have plotted $P(\textbf{r})=P(x)$ on the complex plane of $x$ for $L_z = 0$, $y=0$ and $z=-\infty$.  It resembles a tent with two infinitely high poles at the turning points on the real axis. 

To compare our results with the quantum case, we defined the transmission coefficient $T$ as

\begin{align}\label{TransTEXT}
T = \dfrac{N_{trans}}{N},
\end{align}

\noindent where $N_{trans}$ is the number of transmissions.

We carried out the calculations for the Yukawa and Lennard-Jones interaction potentials. The Yukawa potential is defined as

\begin{equation}\label{yuk}
V_{int}(\mathbf{r}) = -V_{0} \dfrac{r_{0}}{r} e^{-r/r_{0}},
\end{equation}

\noindent where $r = ({\mathbf{x}^{2}+\mathbf{y}^{2}+\mathbf{z}^{2})}^{1/2}$ is the distance separating the two particles in three dimensions and $r_{0}$ is the range of the interaction potential.  And the
Lennard-Jones interaction potential is defined as

\begin{equation}\label{LJp}
V_{int}(\mathbf{r}) = 4 V_0 \bigg[\big(\frac{\sigma}{r}\big)^{12}-\big(\frac{\sigma}{r}\big)^6\bigg],
\end{equation}

\noindent where $V_0$ is the depth of the potential well and $\sigma$ is the van der Waals radius which we set to one.

\subsection{Scattering in the quantum case}\label{ScatteringQuantum}

In the quantum case, we need to solve the Schr\"{o}dinger equation

\begin{equation}\label{SchEq}
\left[-\frac{\hbar^2}{2\mu}\triangledown^2+V(\boldsymbol{r})+\frac{1}{2}\mu\rho^2\omega^2\right]\psi(\boldsymbol{r})=E\psi(\boldsymbol{r}).
\end{equation}

Because of the symmetries of the system, it is more convenient to solve the problem in the spherical coordinates $(r,\theta,\phi)$. Besides, since the angular momentum along the $z$-axis ($L_z=m\hbar$) is conserved, we can separate the $\phi$-variable. Therefore, the problem is reduced to a 2D one. In the asymptotic region where $|z=rcos\theta|\gg R$ ($R$ is the range of the interaction potential), the interaction potential is negligible compared to the confining potential. In this case, the wavefunctions will be of the form $e^{ik_n z}\Phi_{n,m}(\rho,\phi)$ where $\Phi_{n,m}(\rho,\phi)$ is the eigenfunction of the 2D harmonic potential corresponding to the energy $E_{\perp}=\hbar\omega(2n+|m|+1)$, and $k_n=\sqrt{\frac{2\mu}{\hbar^2}E_{||}}$. Then, the asymptotic wave function takes the form ($\rho=r\sin\theta$)

\begin{equation}\label{asymp}
\Psi(r)=e^{ikz}\Phi_{n,m}(\rho,\phi)+\sum_{n'=0}^{n_e} f^{\pm}_{nn'}e^{ik_{n'}|z|}\Phi_{n',m}(\rho,\phi),
\end{equation}

\noindent where ${n_e}$ is the number of open channels, $f^{\pm}_{nn'}$ are the scattering amplitudes that describe the transitions between channels $n$ and $n'$ from which one obtains for the transmission and reflection coefficients

\begin{equation}\label{}
T=\sum_{n'=0}^{n_e}\frac{k_{n'}}{k_n}|\delta_{nn'}+f^{+}_{nn'}|^2,
\end{equation}
and
\begin{equation}\label{R}
R=1-T=\sum_{n'=0}^{n_e}\frac{k_{n'}}{k_n}|f^{-}_{nn'}|^2,
\end{equation}
respectively.
\section{The numerical approach}\label{Numerical}

In this section we describe the numerical methods we used to simulate the scattering of two particles under 2D harmonic confinement in the classical \ref{Classical} and the quantum \ref{Quantum} case.

\subsection{The classical case}\label{Classical}

Our ultimate goal was to calculate the transmission coefficient T of the scattering process for which we had to solve modified Hamilton's equations (Eqs. \ref{ModHamil1}, \ref{ModHamil2}; \ref{Hrel}). Using the $4^{th}$ order Runge-Kutta method led to computational errors near the interaction potential. The \textit{Runge-Kutta-Fehlberg} method, the so-called RK45 \cite{Hoffman} with a global error $\mathcal{O}(h^4)$ allowed us to calculate the errors automatically at each time step during program execution and to modify them suitably; a more efficient algorithm that was stable throughout the calculations, especially near the collision region where highly accurate particle trajectories were obtained.

\begin{figure}[!tbp] 
  \centering
  \includegraphics[width=0.45\textwidth]{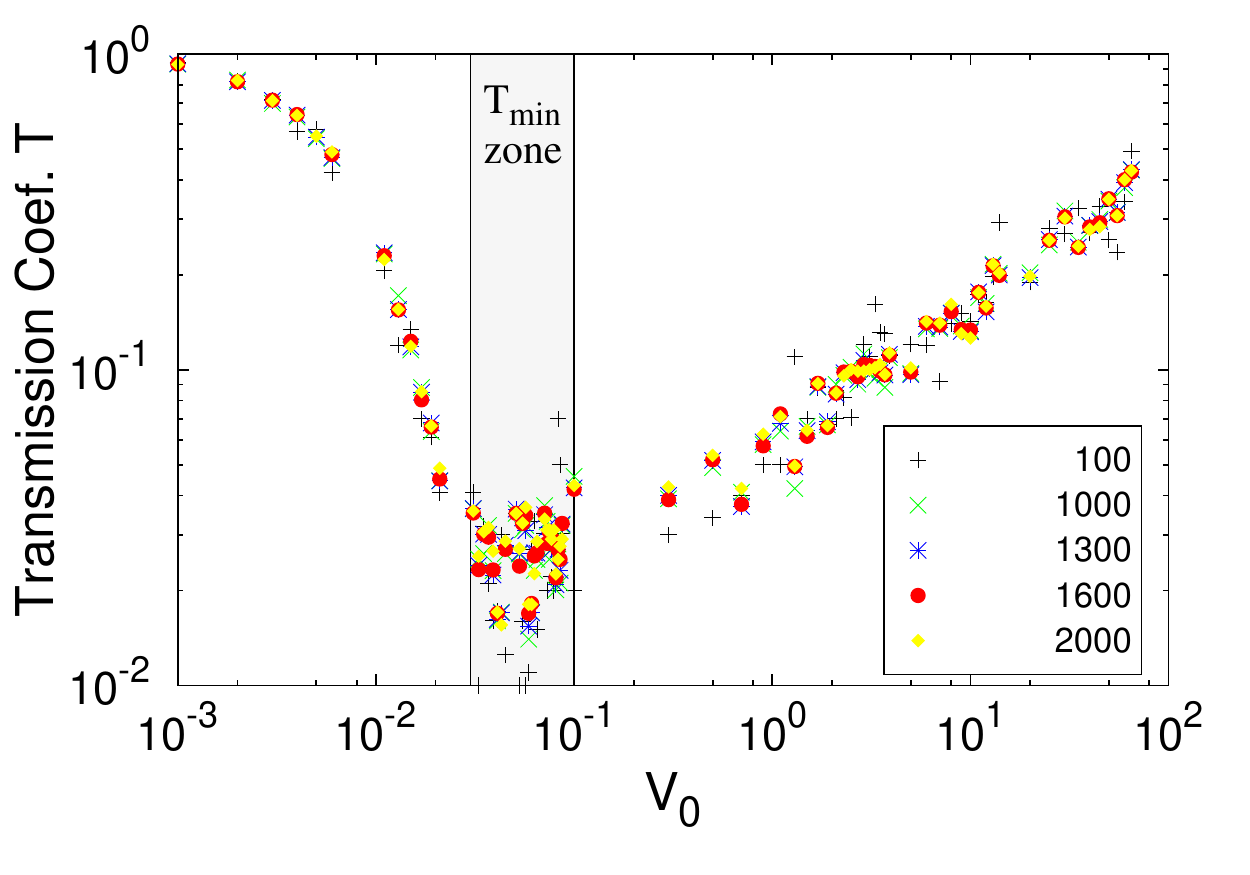}
  \caption{ \footnotesize (Color online) The transmission coefficient T plotted against the interaction potential depth $V_{0}$ for the Yukawa interaction potential for different numbers of events $N$, with $E_{\perp}=1$, $r_{0}=1$, and $L_{z}=0$ in the classical case. All energies are in the units of $\hbar\omega$.}\label{TVNs}
  \end{figure}

Before embarking on extensive calculations, tests were carried out for selected numbers of events from $N = 100$ up to $N = 2000$ by plotting graphs of the transmission coefficient $T$ against the interaction potential depth $V_{0}$. Figure \ref{TVNs} shows the results for $N=100$, $1000$, $1300$, $1600$, and $2000$. It can be seen that there is good convergence of the data points as we increase $N$ above $1000$. 

\begin{figure}[!tbp] 
\centering
\includegraphics[width=0.45\textwidth]{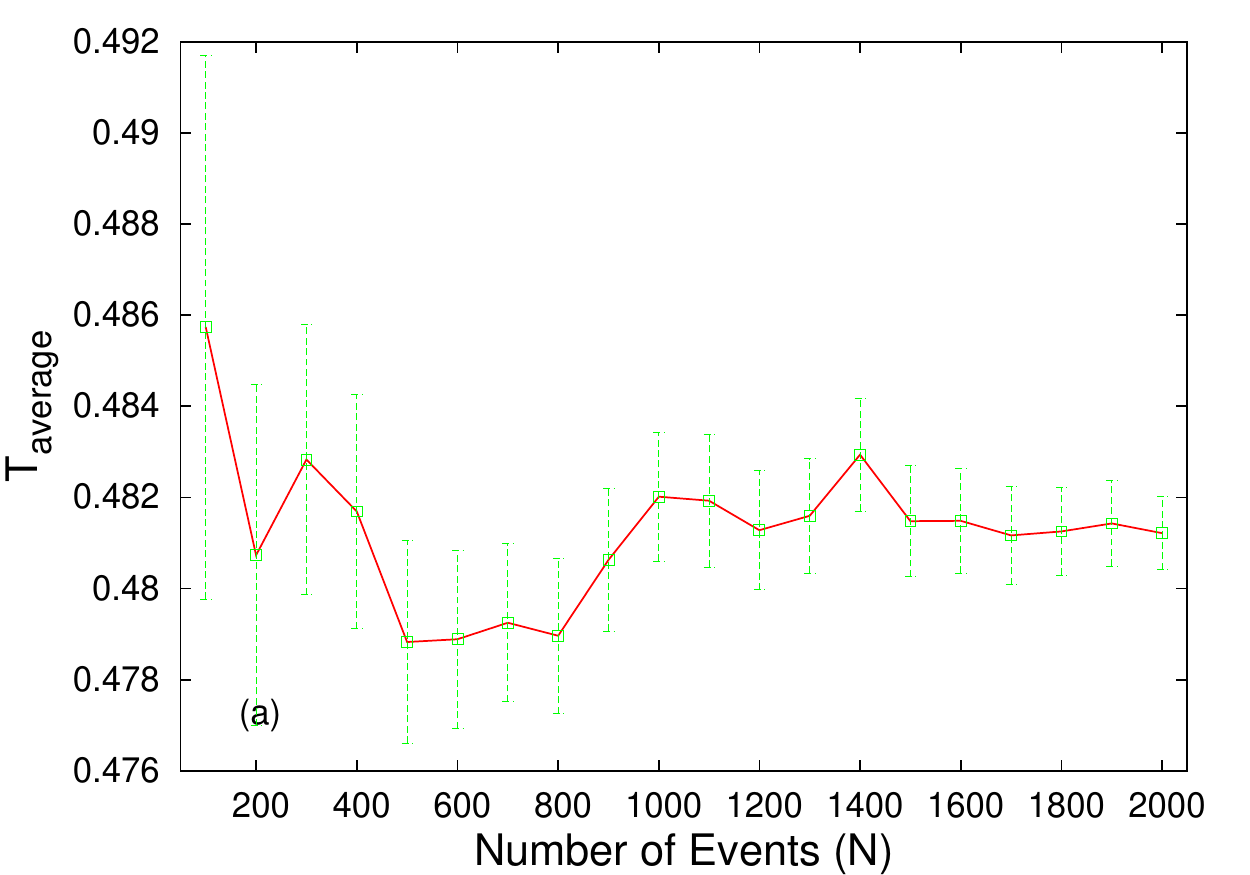} 
\hfill
\includegraphics[width=0.45\textwidth]{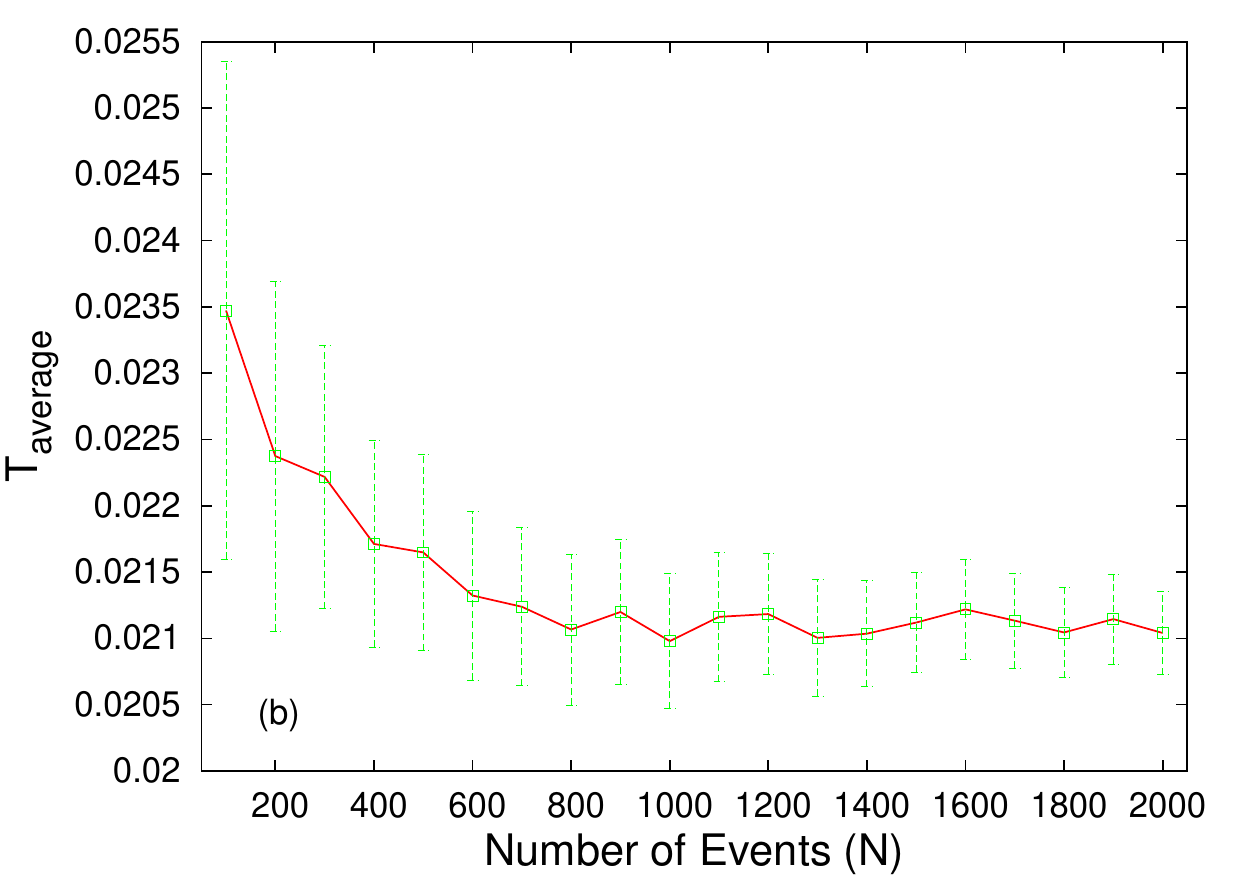} 
\hfill
\includegraphics[width=0.45\textwidth]{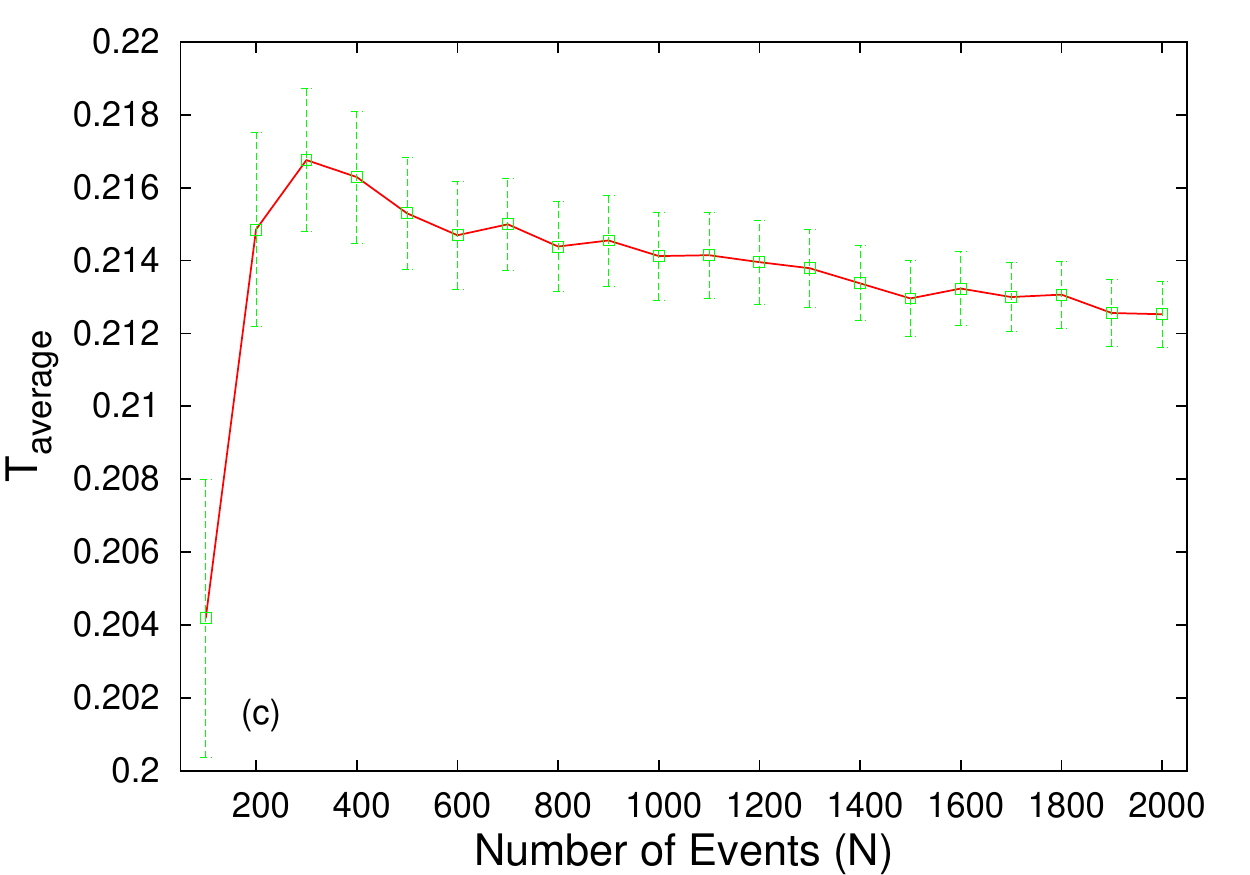} 
\caption{ \footnotesize (Color online) The transmission coefficient $T$ plotted against the number of events $N$ for $V_{0}=0.006$ (a), $V_{0}=0.043$ (b) and $V_0=20.0$ (c) for the Yukawa interaction potential with $E_{\perp}=1$, $r_{0}=1$, and $L_{z}=0$ in the classical case. All energies are in the units of $\hbar\omega$ and each $N$ was repeated $n=100$ times.} \label{3TN}
\end{figure}

To get better statistics, we measured the transmission coefficient as a function of $N$, repeatedly, for three different interaction potential depths $V_0$, one in the interval where $T$ is minimum, what we call the ``$T_{min}$ zone", and two others outside it on the right and the left. The results along with error bars are shown in Fig. \ref{3TN}. The error bars indicate the standard error calculated according to $\Delta T = \sigma / \sqrt{n}$, where $\sigma$ is the standard deviation from the estimated population mean and $n=100$ is the number of times each $N$s was repeated. All three graphs show a clear trend toward the asymptotic values of $T(V_0,N)$ which is achieved for $N \rightarrow \infty$. The error bars are also seen to decrease as $N$ increases. The trends indicated that results obtained for $1600$ events would be conservative enough to strike a balance between statistically acceptable results, given the level of accuracy we required in this work, and computation time.
   
The actual calculations went as follows. For fixed scattering parameters including $V_0$, $L_z$, $E_{\perp}$, and $E_{||}$, we altered the initial scattering conditions by setting the momentum and position of the incident particle to different random values, repeating the process $1600$ times. We further explored the scattering problem by repeating the whole process for different depths of the interaction potential $V_0$. Other physical variables such as $L_z$, $r_0$, $E_{\perp}$, and $E_{||}$ were also changed systematically to obtain the system response in each case. The results are shown in Figures \ref{YukawaMulti5} to \ref{TVEparallelLJ}  where each data point is the result of $1600$ events. The energy used in the calculations had to be kept `real' and constant. The momentum, however, could be taken to be complex.

\subsection{The quantum case}\label{Quantum}

The calculations for the quantum case were based on the discrete-variable numerical method suggested in \cite{Melezhik91} and adapted in \cite{Saeidian08, Saeidian2017}. In this method we discretize the Schr\"{o}dinger equation, transform it into the matrix form, and solve it to obtain the wavefunction from which the transmission coefficient is calculated.

To solve Eq.(\ref{SchEq}) we discretize it on a grid of $\theta$'s and $r$'s. We first discretize the angular variables $\{\theta_j\}^{N_{\theta}}_{j=1}$, $\theta_j$ being the roots of the Legendre's polynomials $P_{N_{\theta}}(\cos\theta)$ and expand the solution $\psi(r, \theta)$ as

\begin{equation}\label{expwave}
\psi(r,\theta)=\sum_{j=1}^{N_{\theta}}\frac{u_j(r)}{r}g_j(\theta),
\end{equation}

\noindent where $g_j(\theta)=\sum_{l=0}^{N_{\theta}-1}P_l(\cos\theta)A_{lj}$. For $\psi(r, \theta)$ to be finite at the origin, we must have

\begin{equation}\label{BouConOrigin}
u_j(r=0,\theta)=0.
\end{equation}

The coefficients $A_{lj}$ are defined as $A_{lj}=[\hat{P}^{-1}]_{lj}$ and $[\hat{P}^{-1}]$ is the inverse of the $N_{\theta}\times N_{\theta}$ matrix $[\hat{P}]$ with elements $[\hat{P}]_{lj}=\lambda_jP_l(\cos\theta_j)$.  $P_l(x)$ are the normalized Legendre Polynomials and $\lambda_j$ are the weights of the Gauss quadrature.  By substituting (\ref{expwave}) into (\ref{SchEq}) we arrive at a system of $N_{\theta}$ coupled equations
\begin{multline}
\big[-\frac{d^2}{dr^2}+2(V(r,\theta_j)+\frac{1}{2}\omega^2 r^2\sin^2\theta_j-E)\big]u_j(r)+\\
\frac{1}{r^2}\sum_{j'=1}^{N_{\theta}}\sum_{l=0}^{N_{\theta}-1}\lambda_j^{\frac{1}{2}}\lambda_{j'}^{\frac{1}{2}}l(l+1)P_l(\cos\theta)A_{lj'}u_{j'}(r)=0.
\end{multline}

By mapping and discretizing $r\in(0, r_m]$ onto the uniform grid $x_j\in(0,1]$ according to $r_j=r_m\frac{e^{\gamma x_j}-1}{e^{\gamma}-1}$, with $j=1, 2, ..., N$ (here $r_m$ is chosen in the asymptotic region $r\rightarrow\infty$, and $\gamma$ is a tuning parameter) and using the finite difference approximation, we solve the above system of equations for a fixed colliding energy $E$ and match the calculated wave function $\psi(r, \theta)$ with the asymptotic behavior (\ref{asymp}) at $r=r_m$ to find the scattering amplitudes $f^{\pm}_{nn'}$.  Knowing $f^+_{nn'}$, one can calculate $T_{nn'}$ from Eq.(\ref{R}).

\section{Results}\label{Results}

Here we present the results of our numerical calculations for a particle moving in a confining 2D harmonic potential scattering off the Yukawa (Subsection \ref{YukawaScattering}) and Lennard-Jones (Subsection \ref{LJScattering}) potentials in order to investigate possible resonances due to confinement using the approaches described in the previous sections.
For simplicity, we set $\mu = 1$, $\hbar = 1$, and $\omega = 1$ unless stated otherwise.

\subsection{Scattering resonance with the Yukawa interaction potential} \label{YukawaScattering}

\begin{figure}[!tbp] 
  \centering
  \includegraphics[width=0.48\textwidth]{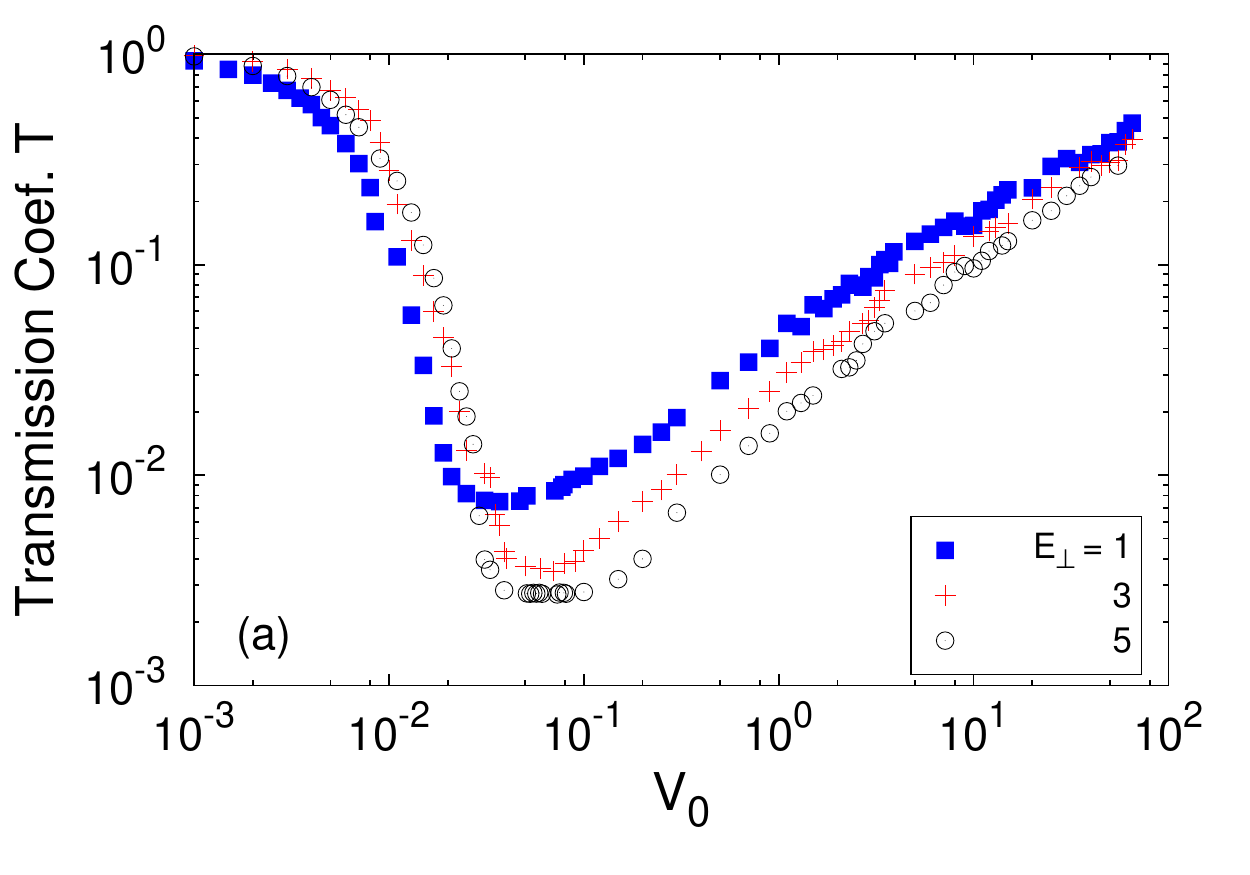}\label{YukawaMulti5Classic}
  \hfill
  \includegraphics[width=0.48\textwidth]{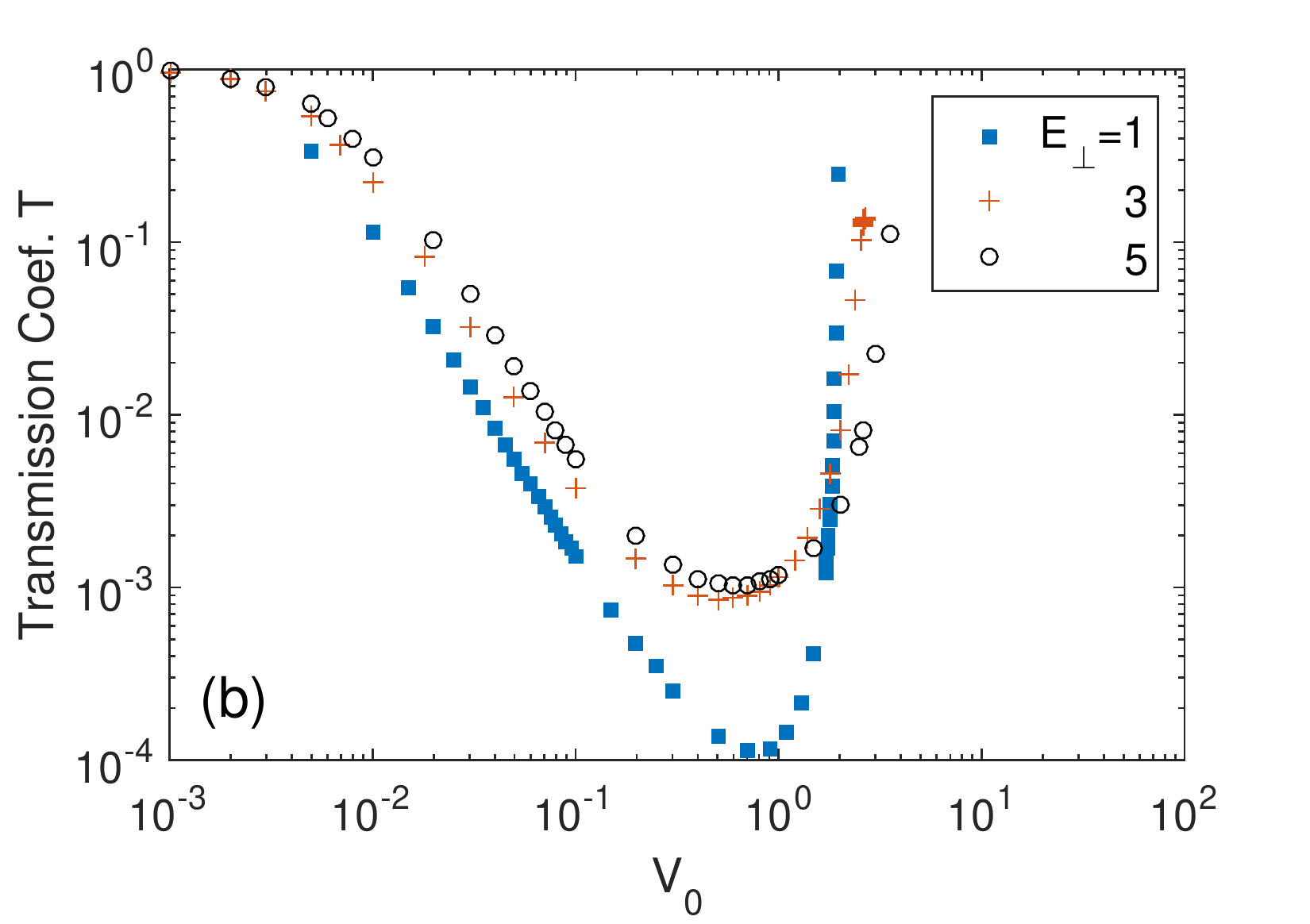}\label{YukawaMulti5Quantum}
  \caption{\footnotesize (Color online) The transmission coefficient $T$ as a function of the potential depth $V_{0}$ for the Yukawa interaction potential with $r_{0}=1$ and $L_{z}=0$, for $E_{||}=10^{-5}$ and different values of the transverse energy $E_{\perp}$ in the classical (a) and quantum (b) cases. All energies are in the units of $\hbar\omega$.}\label{YukawaMulti5}
  \end{figure}

\begin{figure}[!tbp] 
  \centering
\includegraphics[width=0.48\textwidth]{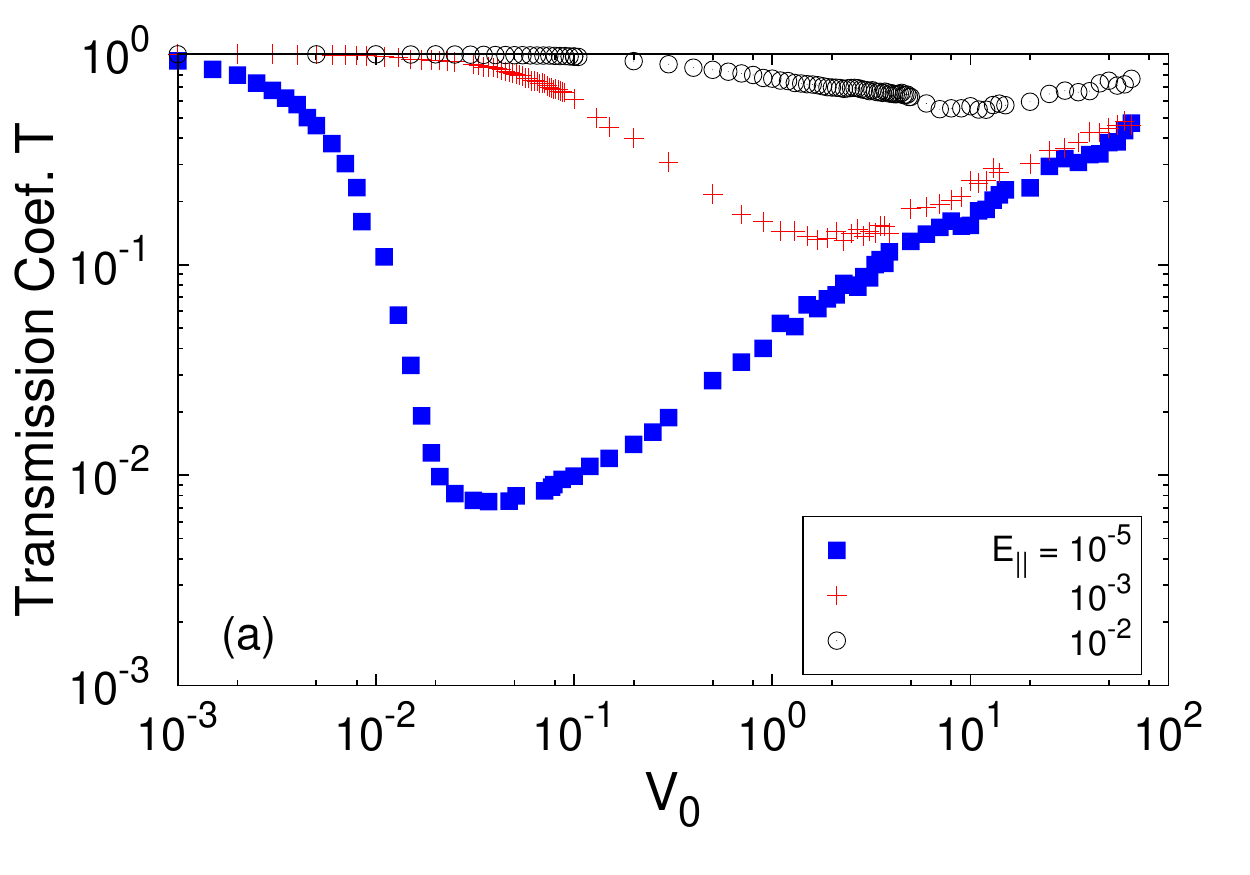} \label{TVEparallela}
  \hfill
\includegraphics[width=0.48\textwidth]{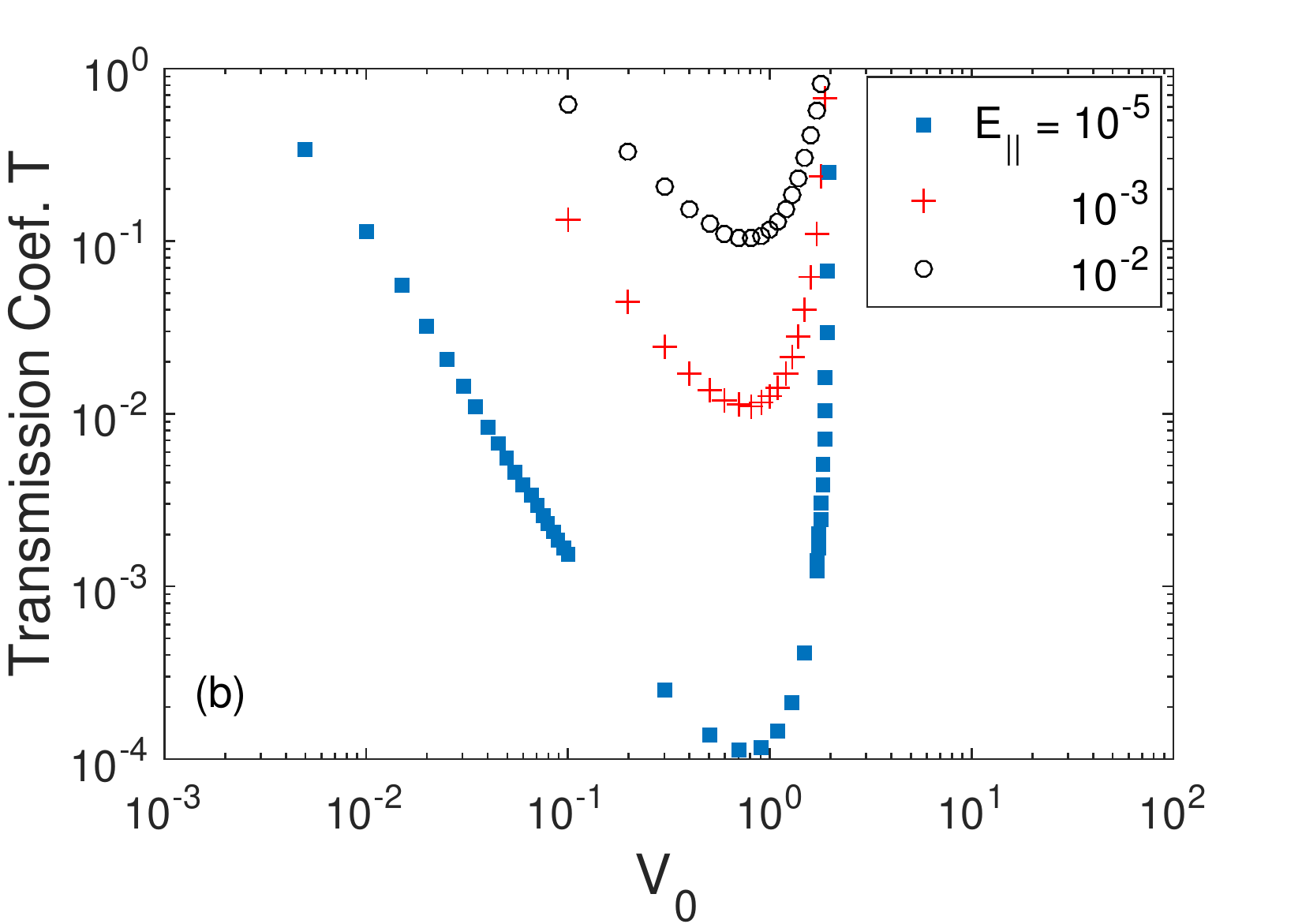} \label{TVEparallelb}
  \caption{ \footnotesize (Color online) The transmission coefficient T plotted against the potential depth $V_{0}$ for the Yukawa interaction potential with $E_{\perp}=1$, $r_{0}=1$, and $L_{z}=0$ and different values of the longitudinal energy $E_{||}$  in the classical (a) and quantum (b) cases. All energies are in the units of $\hbar\omega$.}\label{TVEparallel}
  \end{figure}

Fig. \ref{YukawaMulti5} shows our results for the transmission coefficient $T$ as a function of the interaction potential depth $V_0$ in the classical (a) and quantum (b) cases with $E_{||}=10^{-5}$, $r_{0}=1$, and $L_{z}=0$ for different values of the transverse energy $E_{\perp}$. A comparison of the two graphs shows that as $E_{\perp}$ is increased, in the classical case (a) the minimum of the transmission coefficient $T_{min}$ (corresponding to the occurrence of resonance, CIR) `decreases' and its position shifts to the `right', while in the quantum case (b)  $T_{min}$ `increases', but there is no apparent shift in  its position.
Fig. \ref{TVEparallel} shows the results for a similar set up with $E_{\perp}=1$, $r_{0}=1$, and $L_{z}=0$ for different values of the longitudinal energy $E_{||}$. A comparison of the two graphs shows that as $E_{||}$ is increased, in the classical case (a), $T_{min}$ `increases' and its position shifts to the `right', and in the quantum case (b), $T_{min}$ also `increases' but there is no shift in its position.

  \begin{figure}[h]
  \centering
  \includegraphics[width=0.48\textwidth]{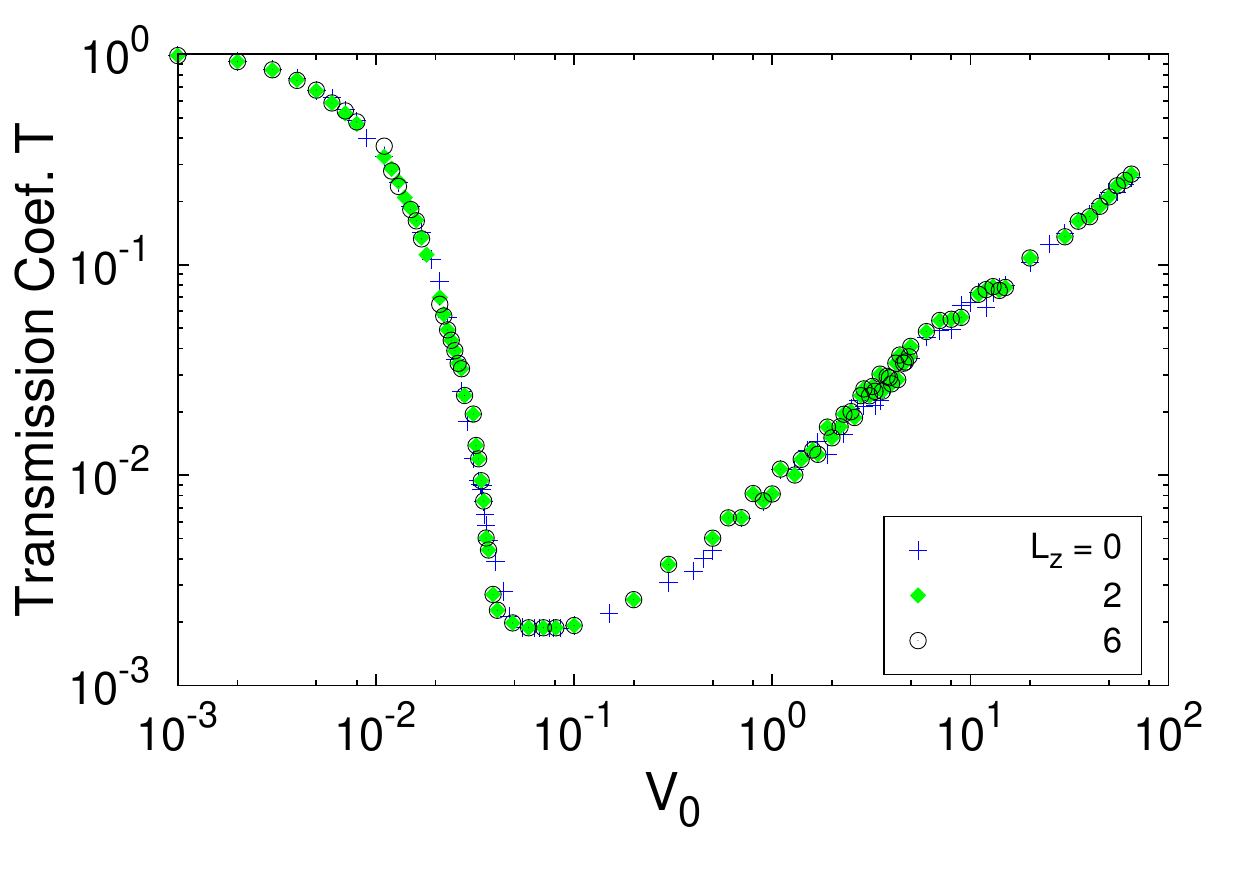}
  \caption{ \footnotesize (Color online) The transmission coefficient T as a function of the potential depth $V_{0}$ for the Yukawa interaction potential with $r_{0}=1$, transverse energy $E_{\perp}=11$, and longitudinal energy $E_{||}=10^{-5}$ for different values of the angular momentum $L_z$. All energies are in the units of $\hbar\omega$.}\label{MultiLz}
  \end{figure}

Figure~\ref{MultiLz} is a plot of the transmission coefficient as a function of $V_{0}$ for different angular momenta ($L_{z}=0, 2,$ and $6$) in the classical case with $E_{\perp} = 11$, $E_{||} = 10^{-5}$, and $r_0 = 1$. Here we have chosen the initial conditions in such a way as to keep $E_{\perp}$ constant. We observe that the occurrence and position of resonance are not significantly affected by the value of the angular momentum, $i.e.$, rotations of the incoming particle around the longitudinal axis. However, to the best of our knowledge, no quantum calculations are available for $L_z \ne 0$ to compare our classical results with except in the case of the zero-range Huang potential where Olshanii has proven analytically that CIR cannot occur for $L_z \ne 0$ \cite{Moore2003}.

  \begin{figure}[!tbp] 
  \centering
\includegraphics[width=0.48\textwidth]{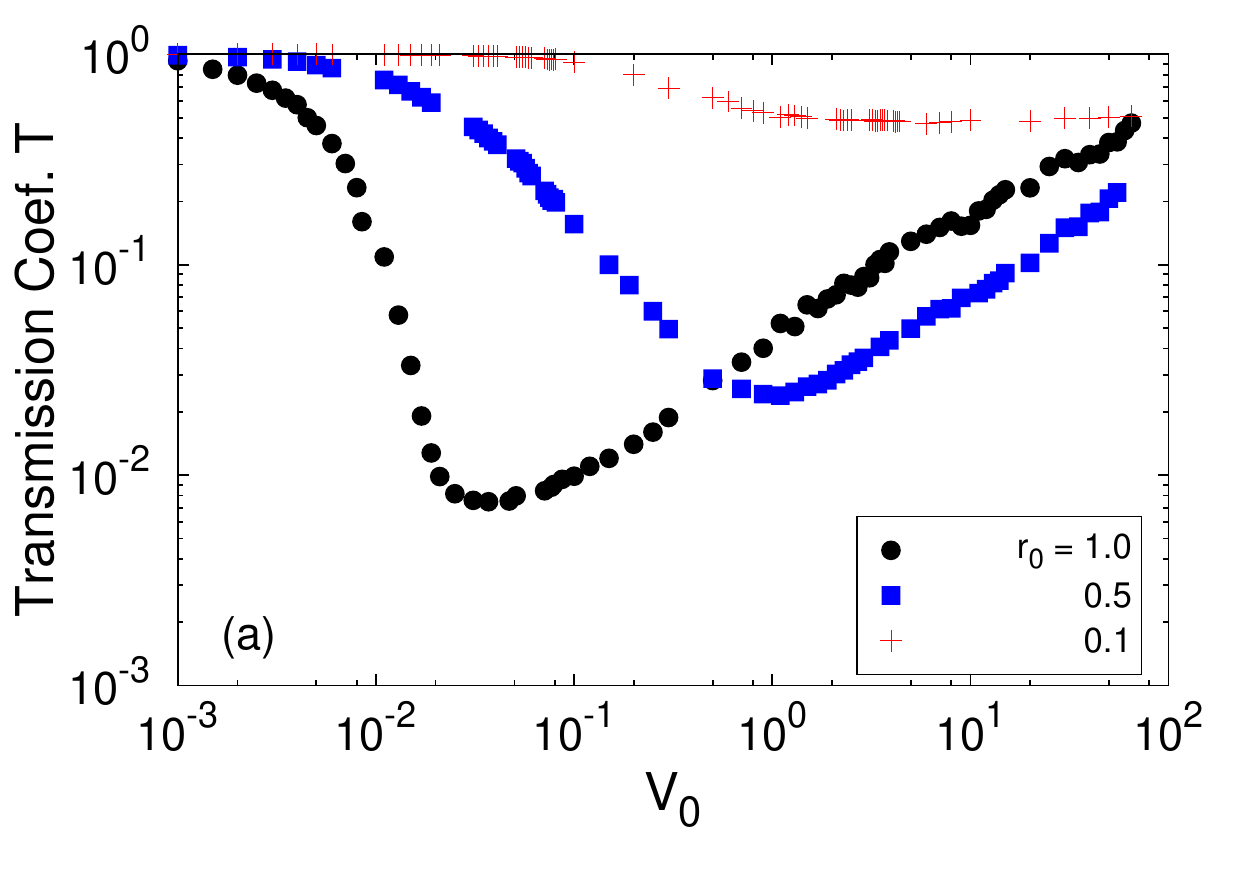} \label{MultiR0a}
  \hfill
\includegraphics[width=0.48\textwidth]{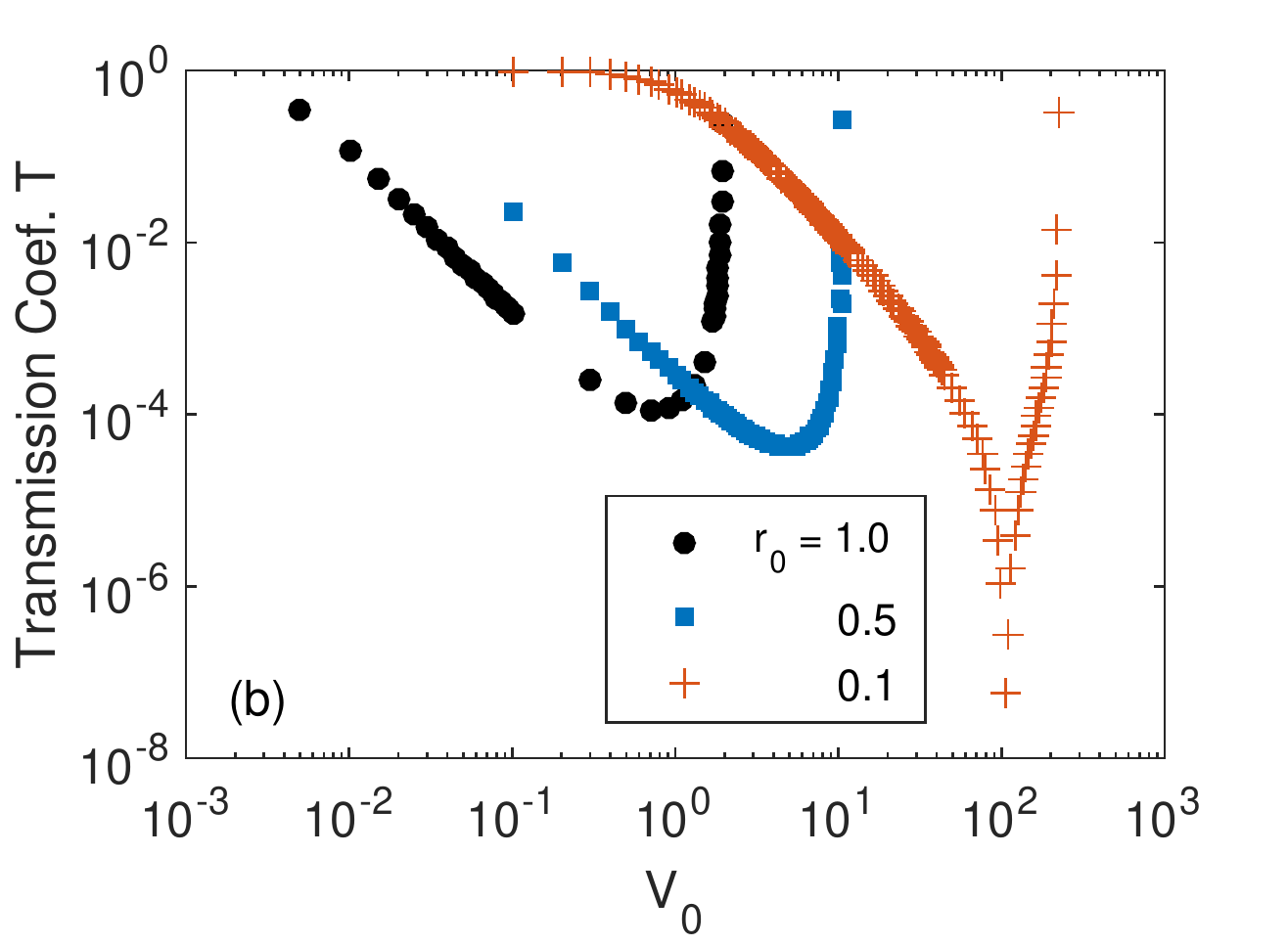} \label{MultiR0b}
  \caption{ \footnotesize (Color online) The transmission coefficient T as a function of the potential depth $V_{0}$ for the Yukawa interaction potential with $L_{z}=0$, transverse energy $E_{\perp}=1$, and longitudinal energy $E_{||}=10^{-5}$ for the classical (a) and quantum (b) cases for different values of the effective range $r_0$. All energies are in the units of $\hbar\omega$.}\label{MultiR0}
  \end{figure}

Up till now, the scattering problem was investigated for the effective range $r_{0}=1$. We now look at the changes in the transmission coefficient $T$ as a function of the potential depth $V_{0}$ for fixed values of the transverse energy $E_{\perp}$, longitudinal energy $E_{||}$, and angular momentum $L_z$ while varying $r_{0}$. The results are plotted in Fig. \ref{MultiR0} for the classical (a) and quantum (b) cases. As can be seen, when $r_0$ is decreased, the ``position" of $T_{min}$ shifts to the right in both classical and quantum cases, but its ``value" increases in the classical case while it decreases in the quantum case.

We should also mention that there are $\it{maxima}$ in the transmission coefficient $T$ both in the classical and quantum case. This occurs when the two atoms do not see each other, $i.e.$, when there is no effective interaction. In the classical case, this corresponds to $T \rightarrow 1$ as $V_0 \rightarrow 0$, where there is no interaction between the particles  (Figs. \ref{YukawaMulti5}(a), \ref{TVEparallel}(a), and \ref{MultiR0}(a)). In the quantum case, this corresponds to the values of $V_0$ for which the scattering length $a_s$ is zero (Figs. \ref{YukawaMulti5}(b), \ref{TVEparallel}(b), and \ref{MultiR0}(b)).

\subsection{Scattering resonance for the Lennard-Jones interaction potential} \label{LJScattering}

In this section, we focus on the resonance that might occur in the Lennard-Jones interaction potential. 

\begin{figure}[!tbp]
  \centering
\includegraphics[width=0.48\textwidth]{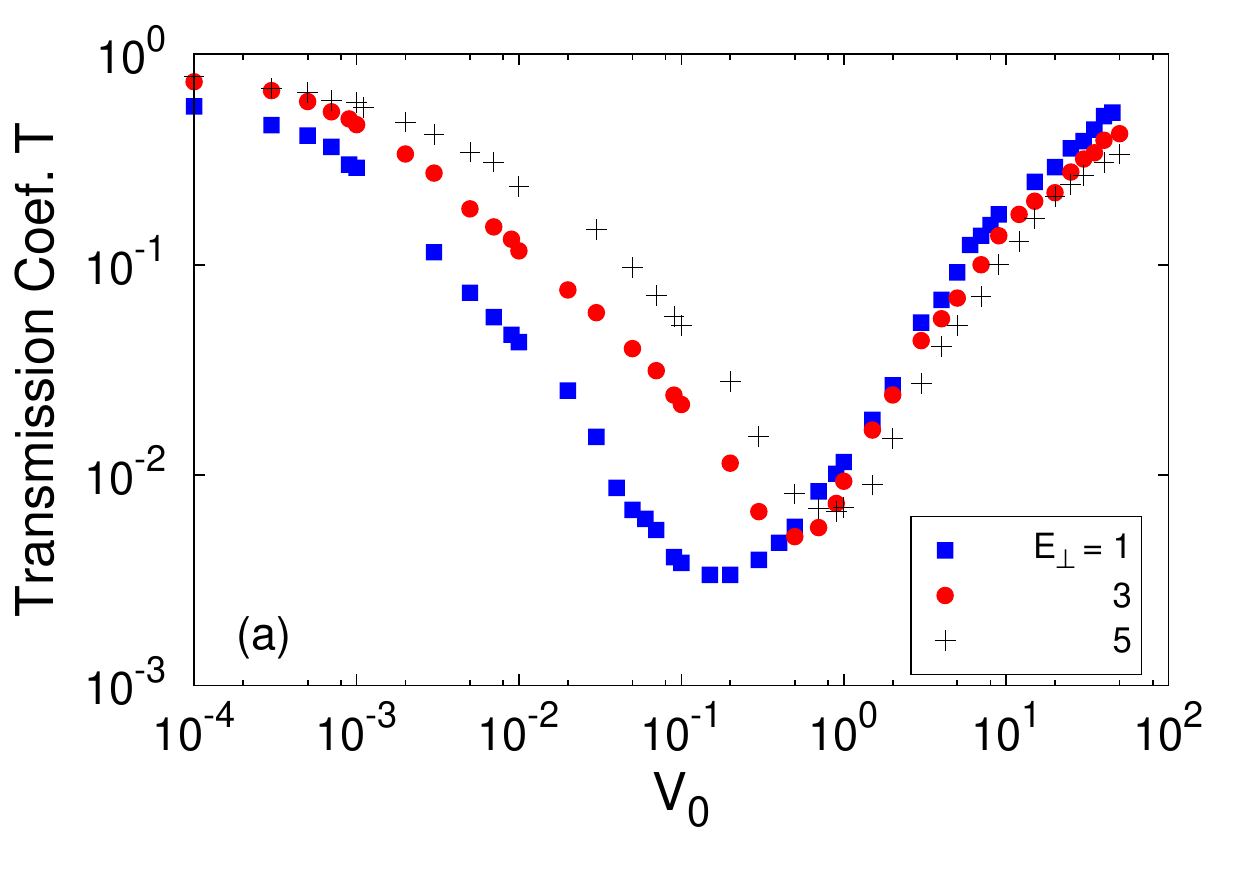}\label{LJMultiClassic}
  \hfill
\includegraphics[width=0.48\textwidth]{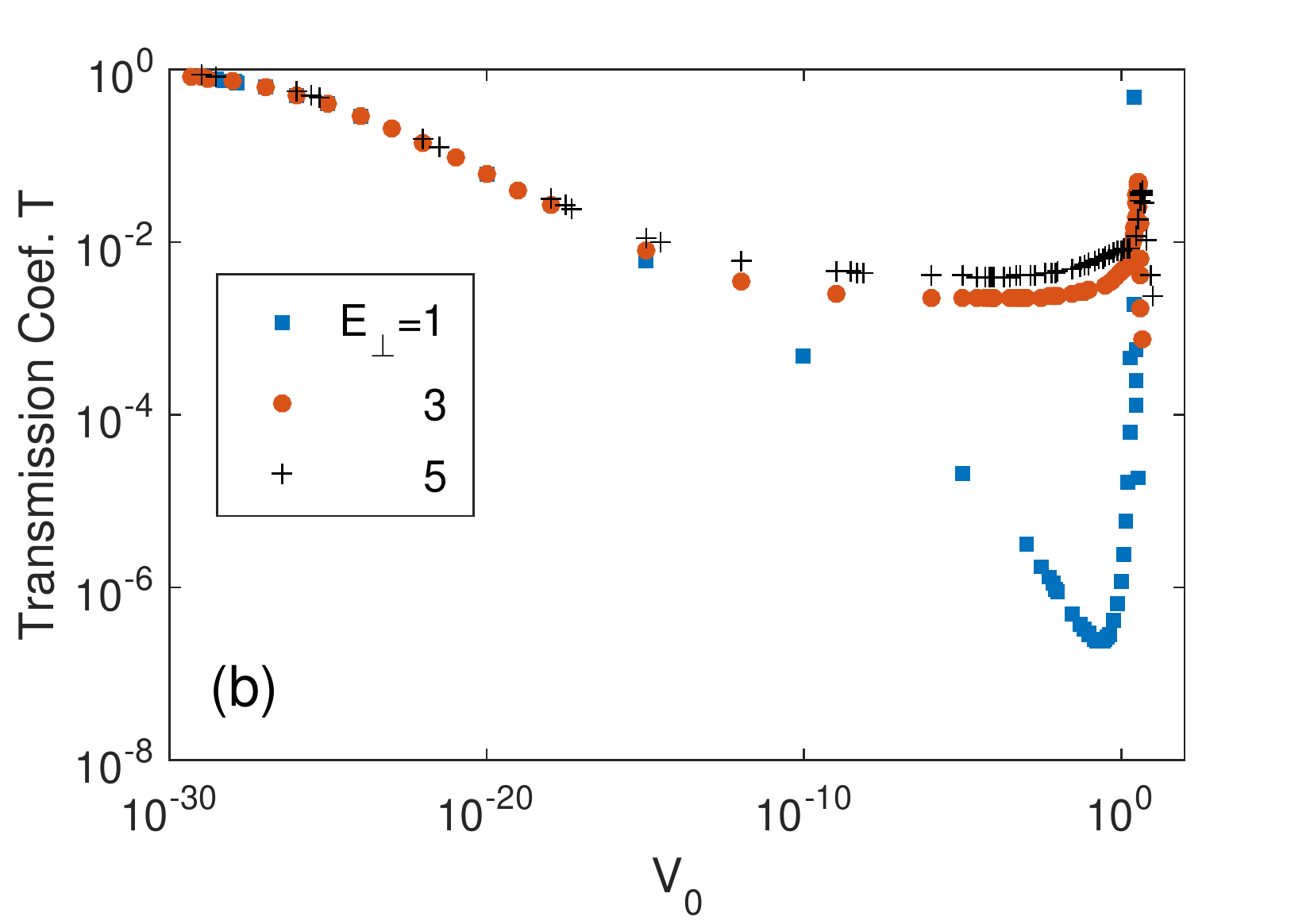}\label{LJMultiQuantum}
  \caption{\footnotesize (Color online) The transmission coefficient T as a function of the potential depth $V_{0}$ for the Lennard-Jones interaction potential with $L_{z}=0$ for $E_{||}=10^{-5}$ and different values of the transverse energy $E_{\perp}$ in the classical (a) and quantum (b) cases. All energies are in the units of $\hbar\omega$.} \label{LJMulti}
  \end{figure}

\begin{figure}[!tbp] 
  \centering
\includegraphics[width=0.5\textwidth]{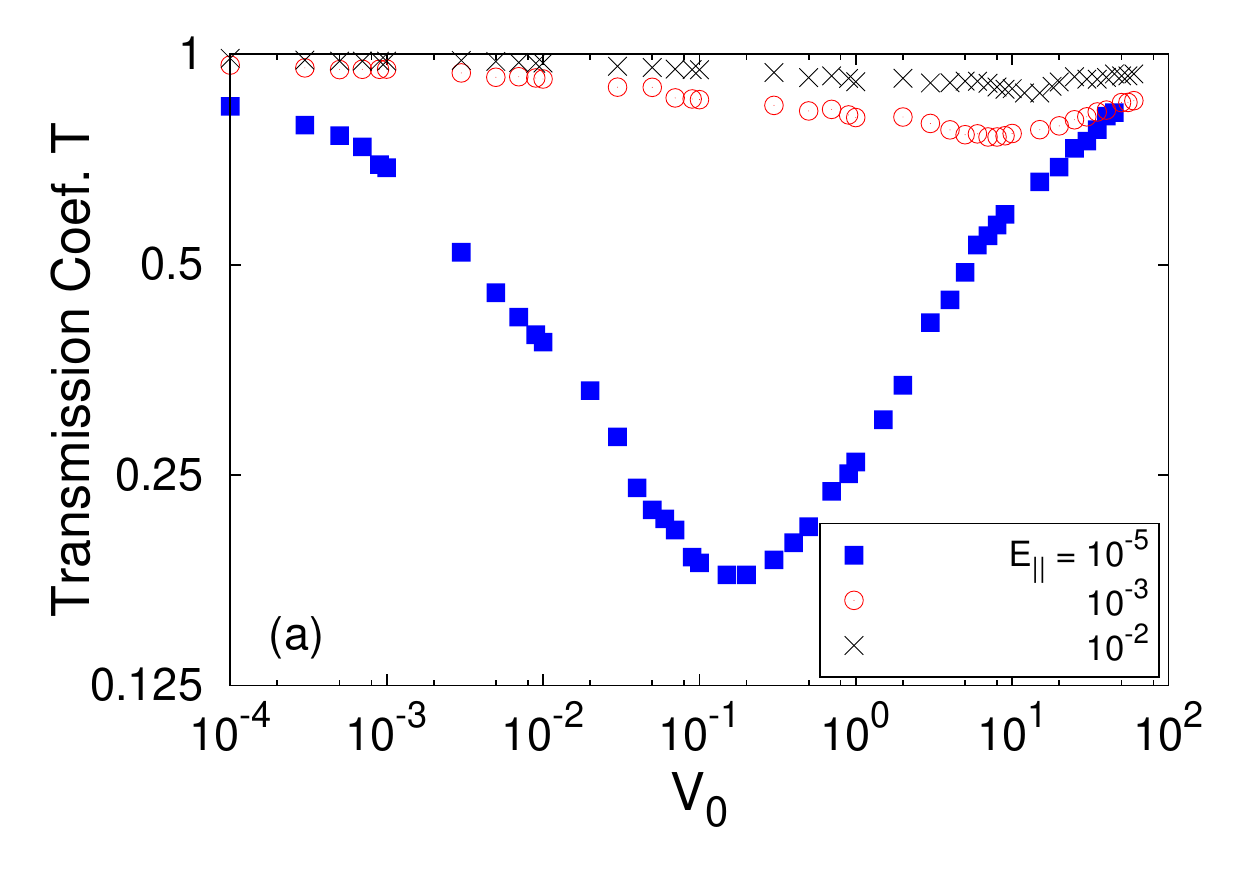} \label{TVEparallelLJa}
  \hfill
\includegraphics[width=0.5\textwidth]{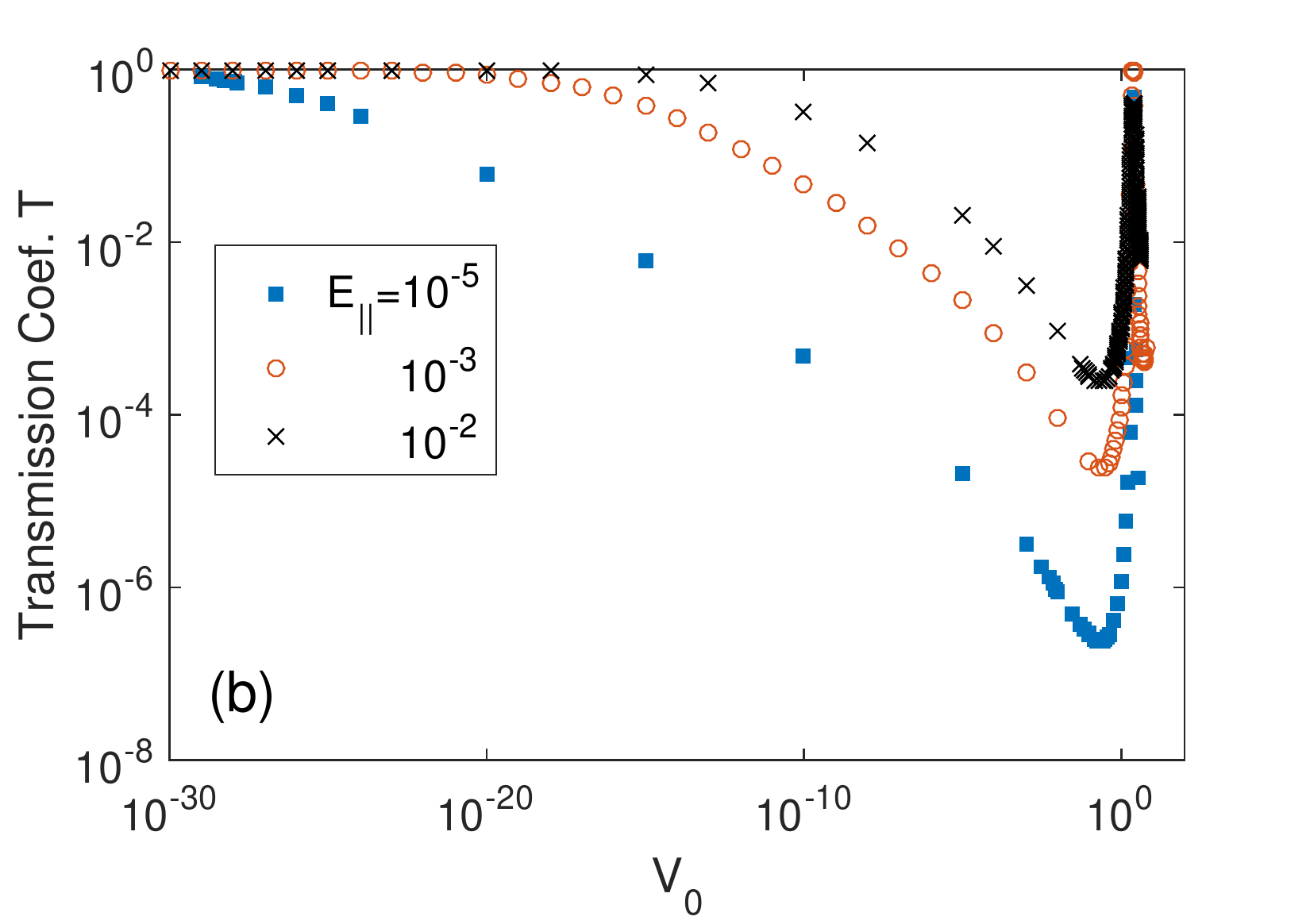} \label{TVEparallelLJb}
  \caption{ \footnotesize (Color online) The transmission coefficient T vs. the potential depth $V_{0}$ of the Lennard-Jones interaction potential with $E_{\perp}=1$ and $L_{z}=0$ and different values of the longitudinal energy $E_{||}$ in the classical (a) and quantum (b) cases. All energies are in the units of $\hbar\omega$.}\label{TVEparallelLJ}
  \end{figure}

Fig. \ref{LJMulti} shows our results for the transmission coefficient $T$ as a function of the interaction potential depth $V_0$ in the classical (a) and the quantum (b) case with $E_{||}=10^{-5}$ and $L_{z}=0$ for different values of the transverse energy $E_{\perp}$. A comparison of the two graphs shows that as $E_{\perp}$ is increased, in the classical case (a), $T_{min}$ `increases' and its position shifts to the `right', and in the quantum case (b),  $T_{min}$ also `increases' but its position shifts to the `left'.  Fig. \ref{TVEparallelLJ} shows the results for a similar set up with $E_{\perp}=1$ and $L_{z}=0$ for different values of the longitudinal energy $E_{||}$. A comparison of the two graphs shows that as $E_{||}$ is increased, in the classical case (a), $T_{min}$ `increases' and its position shifts to the `right', and in the quantum case (b), $T_{min}$ also `increases' but there is no shift in its position.
   
As in the case of the Yukawa interaction potential  (subsection A), here too there are $\it{maxima}$ ($T\rightarrow 1$) in the transmission coefficient $T$ both in the classical and quantum cases when there is no effective interaction between the atoms. Figs. \ref{LJMulti}(a) and \ref{TVEparallelLJ}(a) show this in the classical case and Figs. \ref{LJMulti}(b) and \ref{TVEparallelLJ}(b) show it in the quantum case.

\section{Discussion} \label{Discussion}
In this section, we will put forward arguments to explain the observations in Figs. \ref{YukawaMulti5}-\ref{TVEparallelLJ} and the apparent discrepancies, if any, in the response of the system to changes in system parameters both in the classical and quantum regimes. As pointed out in Subsection VI. A., the CIR that occurs in the classical case is similar to the quantum case in origin (formation of a temporary molecular state) and type (the Feshbach resonance). However, since they act in different realms (quantum mechanics is non-local and probabilistic while classical mechanics is localized and deterministic), we should not expect their CIRs to occur at similar $V_0$s or for the spread of their values to be similar . Also, the ``positions" of the classical and quantum $T_{min}$s might even differ by one or two orders of magnitude, as may be seen in the $T-V_0$ graphs.  In addition, in the classical case, the incident particle is more likely to be found away from the trap axis, seeing, on average, less of the interaction potential, leading to a weaker binding (a less stable quasi-molecular state). In the quantum case, however, the incident particle is more likely to be found near the axis, seeing, on average, more of the interaction potential, leading to a stronger binding and a more stable quasi-molecular state. Therefore, the classical results for $T_{min}$ are expected to have higher ``values" as shown in the $T-V_0$ graphs.

Table \ref{T1} shows the behavior of the transmission coefficient minimum $T_{min}$ in the $T-V_0$ graphs (Figs. \ref{YukawaMulti5}, \ref{TVEparallel}, \ref{LJMulti}, and \ref{TVEparallelLJ}) as the transverse energy $E_{\perp}$ or longitudinal energy $E_{||}$ of the incoming particle is increased. We have indicated the response of the system, $i.e.$, an increase or decrease in the value of $T_{min}$, by up and down arrows ($T_{min} \uparrow\ $ or $\downarrow\ $) and a shift in its position to the right or to the left, by right and left arrows ($T_{min} \rightarrow\ $ or $\leftarrow\ $) in $T-V_0$ graphs.  A shift to the right (left) in the $T_{min}$ position implies that a deeper (shallower) $V_0$ is employed to scatter the approaching particle. 

According to this table, in the classical case, as the transverse energy $E_{\perp}$ increases, $T_{min}$ `decreases' under the Yukawa potential ($T_{min} \downarrow$) and `increases' under the Lennard-Jones potential ($T_{min} \uparrow$) while the shift in its position is always to the `right' ($T_{min} \rightarrow$) under either potential. On the other hand, in the quantum case, as the transverse energy $E_{\perp}$ increases, the changes in $T_{min}$ are the same under either potential, $i.e.$, $T_{min}$ `increases' ($T_{min} \uparrow$) while the shift in its position is always to the left ($T_{min} \leftarrow$).  However for the Yukawa potential this is too small to be significant.  

As for the changes in the longitudinal energy of the particle $E_{||}$, Table \ref{T1} shows that in the classical case, as $E_{||}$ increases, $T_{min}$ increases and shifts to the right ($T_{min} \uparrow\rightarrow$) under the influence of either interacting potential, while in the quantum case, increasing $E_{||}$ results in an increase in $T_{min}$ with no shift in its position ($T_{min} \uparrow\times$) under either potential.

In addition to considering the changes in the energy of the system, we will discuss the physical reasons behind the response of the system to the changes in the angular momentum $L_z$ and the potential range $r_0$ for the Yukawa potential.

In the following subsections we will first present a physical interpretation of the CIR phenomenon and then discuss the scattering response of the system in the classical and quantum cases, respectively.

\begin{table}[!htbp]
\caption{The classification of the behavior of the transmission coefficient minimum $T_{min}$ as observed in the $T-V_0$ graphs of Figs. \ref{YukawaMulti5}, \ref{TVEparallel}, \ref{LJMulti}, and \ref{TVEparallelLJ} when the transverse energy $E_{\perp}$ or longitudinal energy $E_{||}$ are increased. }
\begin{center}
\renewcommand{\arraystretch}{1.2}
\begin{ruledtabular}
\begin{tabular}{@{\,}l*{5}{c}@{\,}}
 &&& \multicolumn{2}{c}{\thead{$T_{min}$}}\\
\cmidrule[\heavyrulewidth]{4-5}
&&& \thead{Classical} & \thead{Quantum} \\
\hline
    \multirow{2}{*}{Yukawa} & Fig. \ref{YukawaMulti5} & $E_{\perp} \uparrow$ & $\downarrow\rightarrow$ & $ \uparrow\times$\\
    \cmidrule{2-5}                & Fig. \ref{TVEparallel} & $ E_{||} \uparrow$      & $\uparrow\rightarrow$      & $\uparrow\times$\\
\hline
    \multirow{2}{*}{Lennard-Jones} & Fig. \ref{LJMulti} & $E_{\perp} \uparrow$ & $\uparrow\rightarrow$ & $\uparrow\leftarrow$\\
    \cmidrule{2-5}                            & Fig. \ref{TVEparallelLJ} & $E_{||} \uparrow$       & $\uparrow\rightarrow$ & $\uparrow\times$\\
\end{tabular}
\end{ruledtabular}
\end{center}
\label{T1}
\end{table}

\subsection{ Physical interpretation of CIR}

In the $\bold{quantum}$ $ \bold{case}$, CIR occurs when the energy of the binding state of the interaction potential corresponding to the closed channel is near the energy threshold of the open channel. In this case, the system may temporarily enter a molecular state. After emerging from this, it is either transmitted or reflected. In other words, near CIR, the interaction potential couples the excited eigenstate of the confining potential corresponding to the closed channel, with the eigenstate(s) of the open channel(s). The resulting phenomenon is a Feshbach resonance.

\begin{figure}[!tbp] 
  \centering
\includegraphics[width=0.5\textwidth]{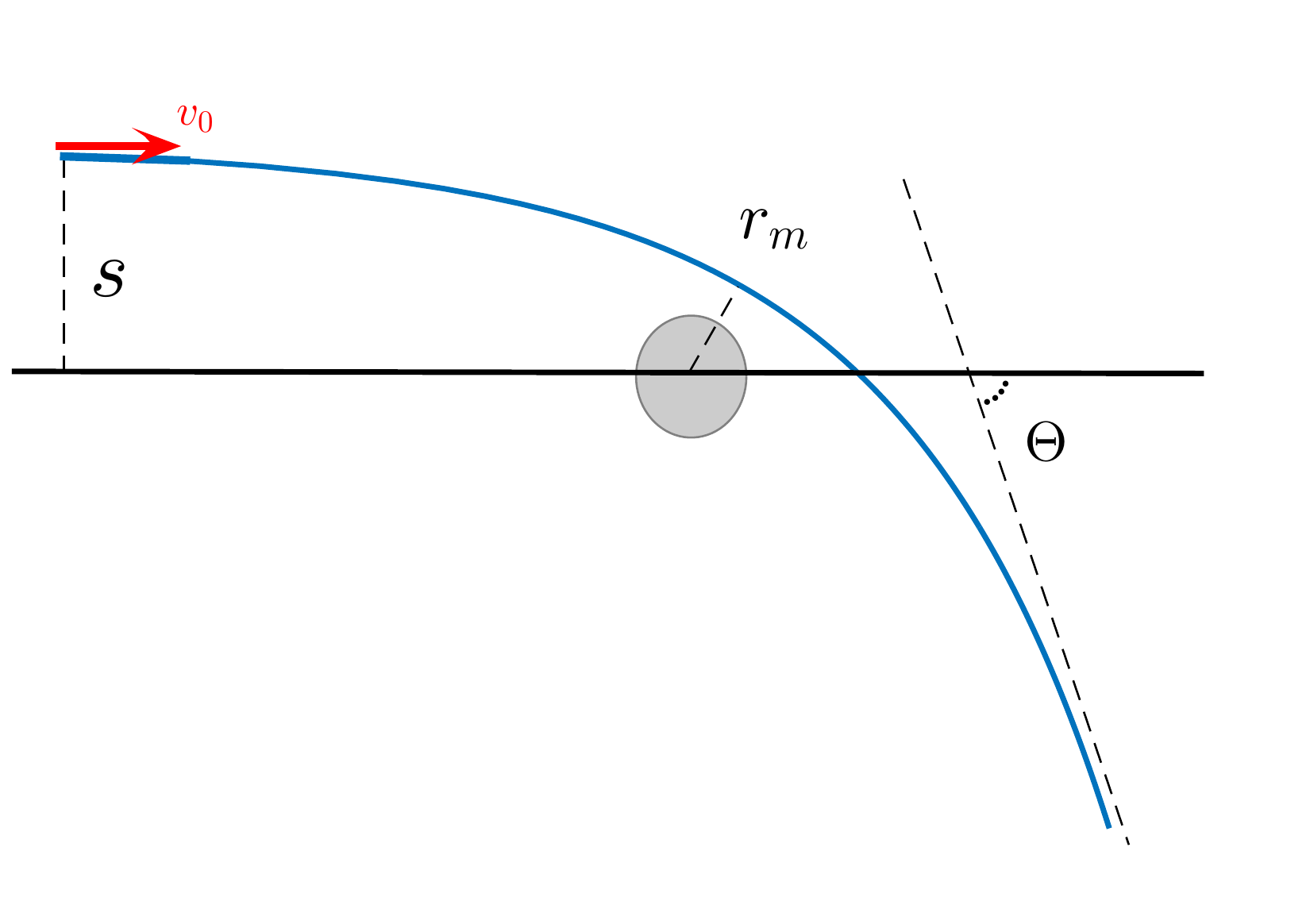}
  \hfill
  \caption{ \footnotesize (Color online) Schematic drawing  of the orbit of an incident particle with initial velocity $v_0$ and impact parameter $s$, scattered  by a centre of force.}\label{3Dorbit}
  \end{figure}

To consider CIR in a classical setting, we will first look at the scattering of a particle with mass $\mu = 1$ and incident energy $E$, from a short-range central potential $V(r)$ in $\bold{free}$ $\bold{ space}$  ($i.e.$, without a confining potential). When the particle approaches the centre of the force, its path curves away from the original straight line (Fig. \ref{3Dorbit}). After the particle moves away from the centre of the potential, the force exerted on the particle gradually diminishes and the particle  continues on a straight line. The scattering angle between the incident and scattered directions is obtained from the following relationship

\begin{equation}\label{ScattAngle}
\Theta=-\pi+2\int_{r_m}^{\infty}\frac{J/r^2}{\sqrt{2[E-V(r)-J^2/2r^2]}}dr
\end{equation}
where $r_m$ is the distance of the closest approach and $J=sv_0$.  The impact parameter $s$ is defined as the perpendicular distance between the centre of force and the incident velocity, while $v_0=\sqrt{2E}$. 

\begin{figure}[!tbp] 
  \centering
\includegraphics[width=0.5\textwidth]{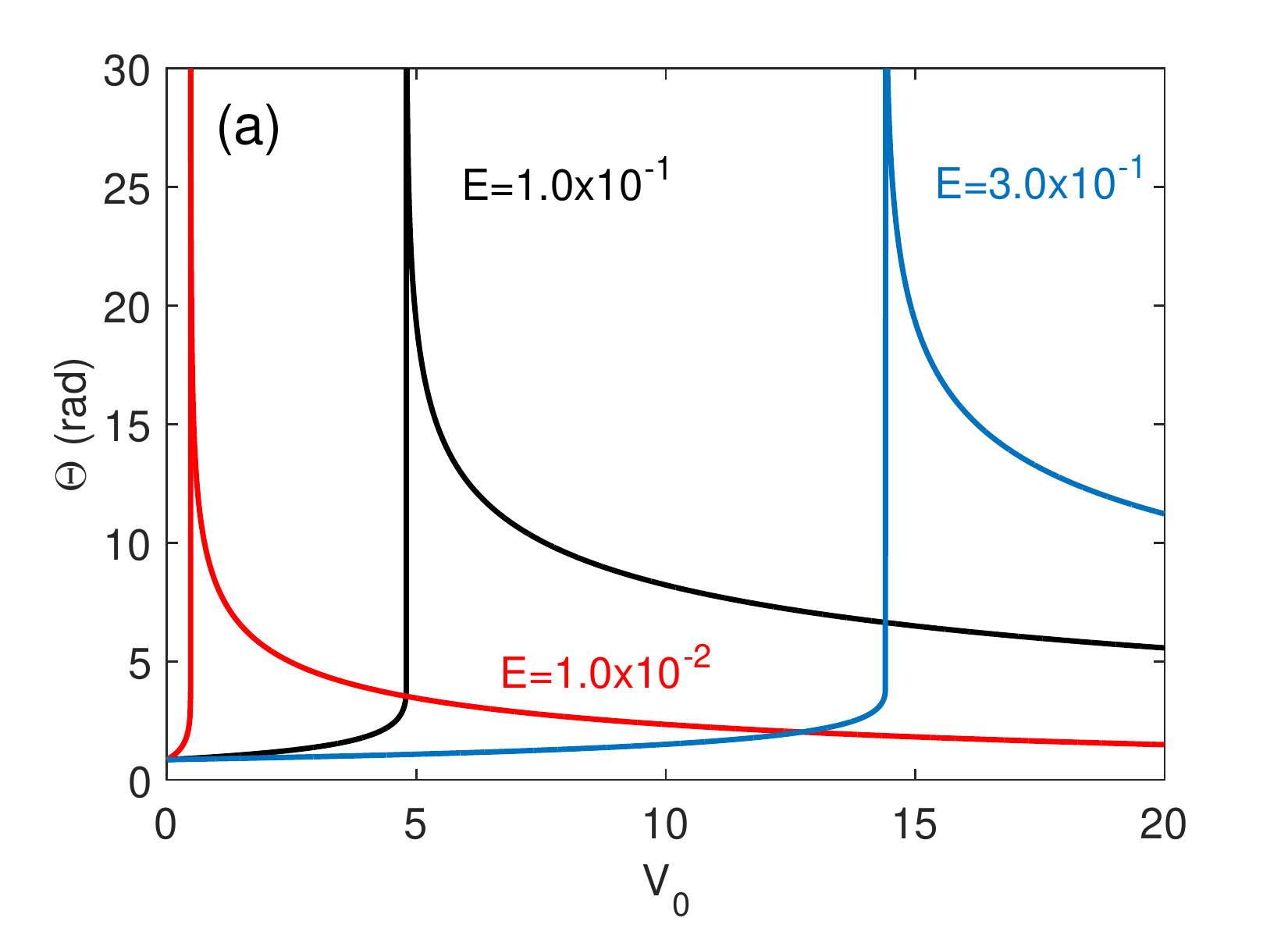} \label{orbit}
  \hfill
\includegraphics[width=0.5\textwidth]{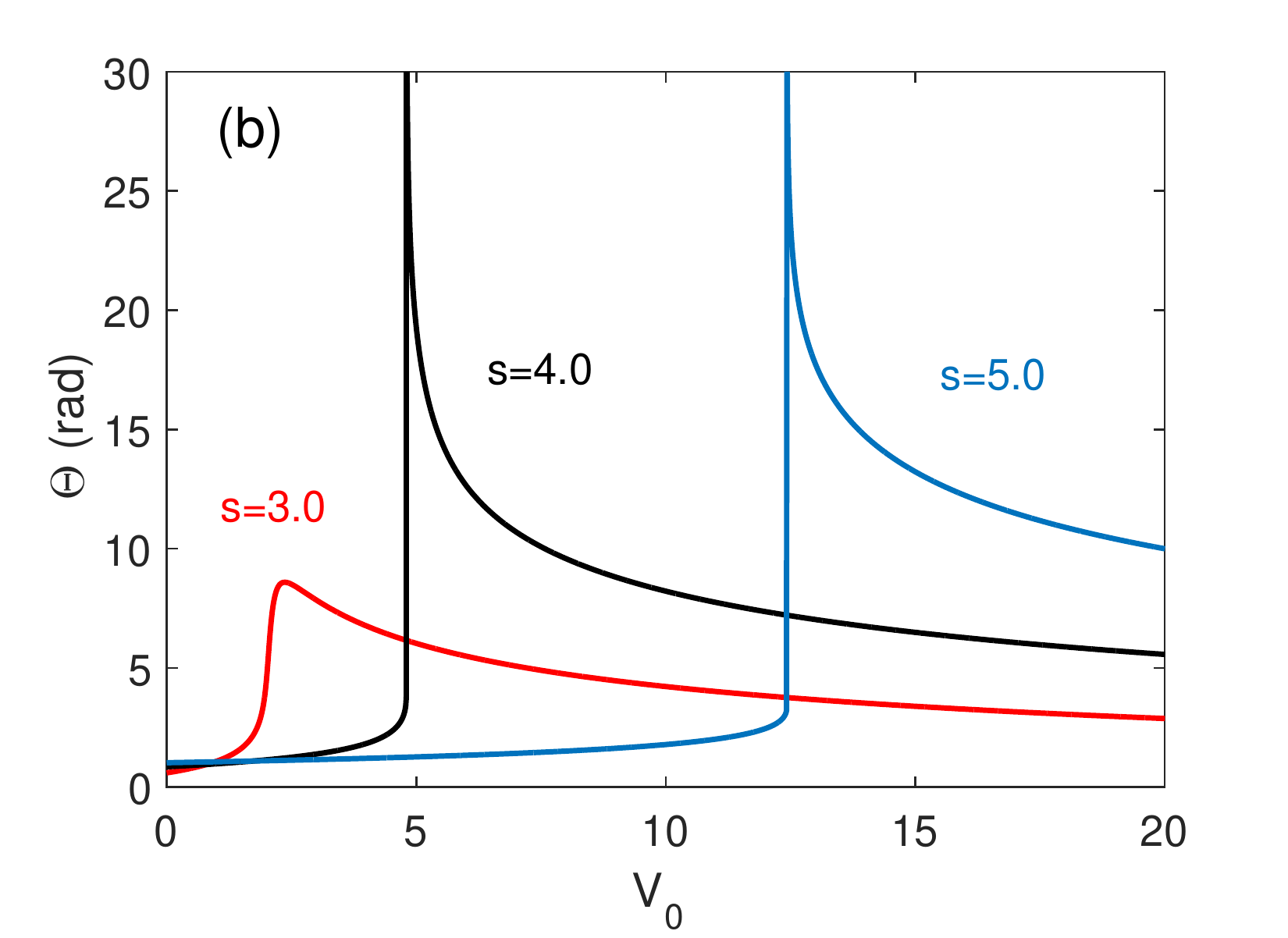} \label{orbit}
  \caption{ \footnotesize (Color online) The scattering angle $\Theta$ versus the potential depth $V_0$ for the Yukawa potential for the impact parameter $s=4$ for different values of the incident energy $E$ (a) and for $E=1.0\times 10^{-1}$ for different values of $s$ (b).}\label{ThetaV0}
  \end{figure}

\begin{figure}[!tbp] 
  \centering
\includegraphics[width=0.5\textwidth]{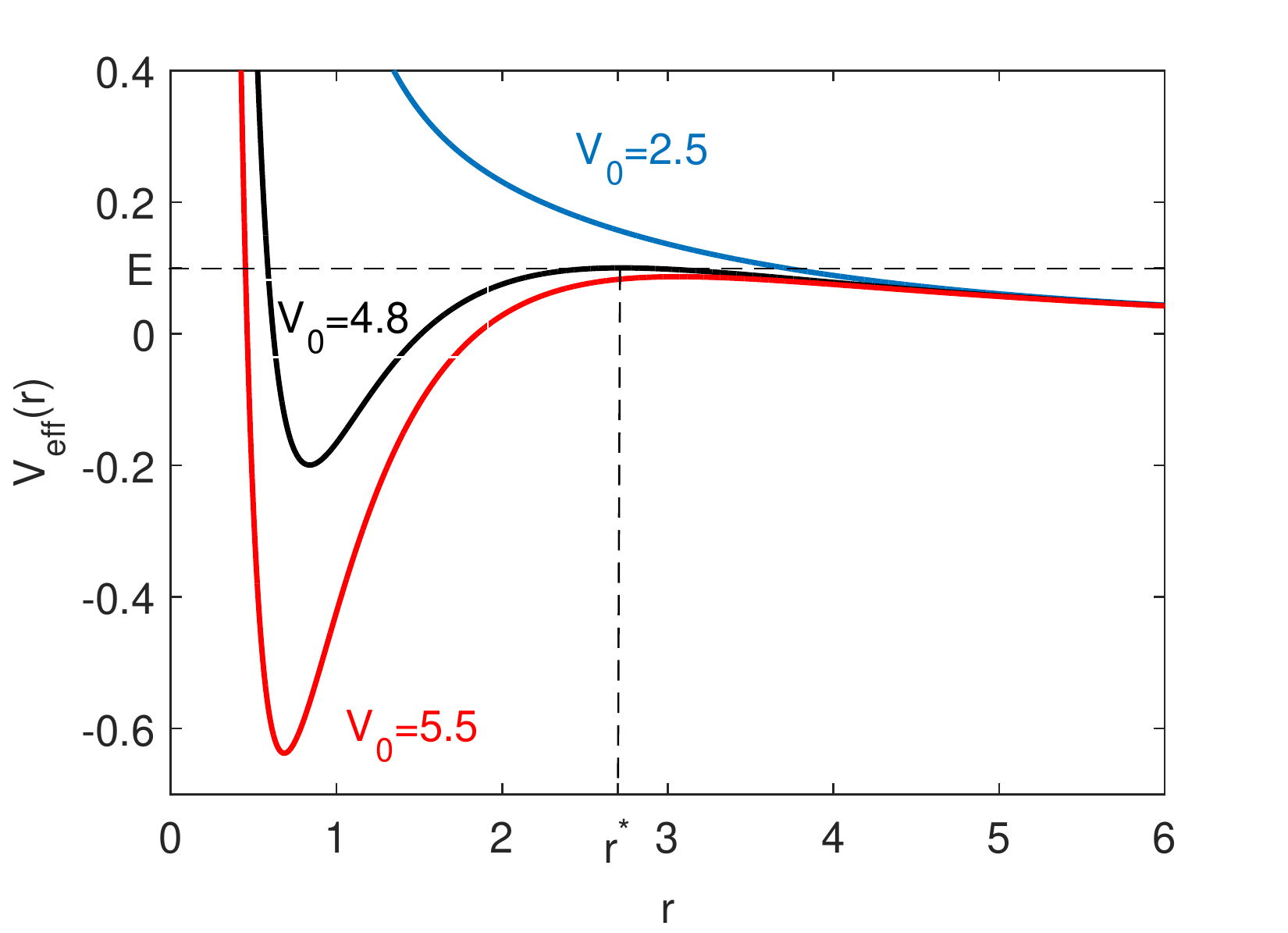} 
  \hfill
  \caption{ \footnotesize (Color online) The effective potential  (\ref{VeffPot}) versus the inter-particle distance $r$ for $s=4.0$ and $E=1.0\times 10^{-1}$ for the Yukawa interaction potential at different values of the potential depth $V_0$.}\label{effPot}
  \end{figure}

$\Theta$ is a function of $s$, $V_0$, and $E$. $\Theta > 2 \pi$ implies that the particle revolves around the centre of force a number of times before scattering. Fig.\ref{ThetaV0} shows plots of $\Theta$ against $V_0$ for different values of the scattering parameters for the Yukawa potential. As can be seen, resonance occurs for certain values of the scattering parameters signaled by the divergence in $\Theta$.  To see this, we need to consider the equation of motion for the r-component
\begin{equation}\label{requ}
\ddot r=-\frac{d}{dr}V_{eff}(r)
\end{equation}
where the effective potential $V_{eff}$ is defined as
\begin{equation}\label{VeffPot} 
V_{eff}(r)=V(r)+\frac{J^2}{2r^2}.
\end{equation} 
Fig.\ref{effPot} is a graph of $V_{eff}(r)$ against the inter-particle distance $r$ for the impact parameter $s=4$ and the incident energy $E =1.0\times 10^{-1}$ for the Yukawa potential with different values of $V_0$. Since the repulsive centrifugal barrier dominates at large $r$, the equivalent potential for sufficiently large $V_0$ will exhibit a bump. For a special value of $V_0$ (say $V_0=4.8$ here) when an incident particle with impact parameter $s$ and energy $E$ reaches the location $r^*$ where $V_{eff}$ is maximum, the radial velocity $\dot r$ is zero. In addition, $\ddot r$ is also zero, $i.e.$, the particle is in an unstable equilibrium state in the radial direction. In the absence of any perturbation, the particle having reached $r^*$ would circle around the centre of force indefinitely at the same distance without ever escaping! \cite{Goldstein}.

\begin{figure}[!tbp] 
  \centering
\includegraphics[width=0.5\textwidth]{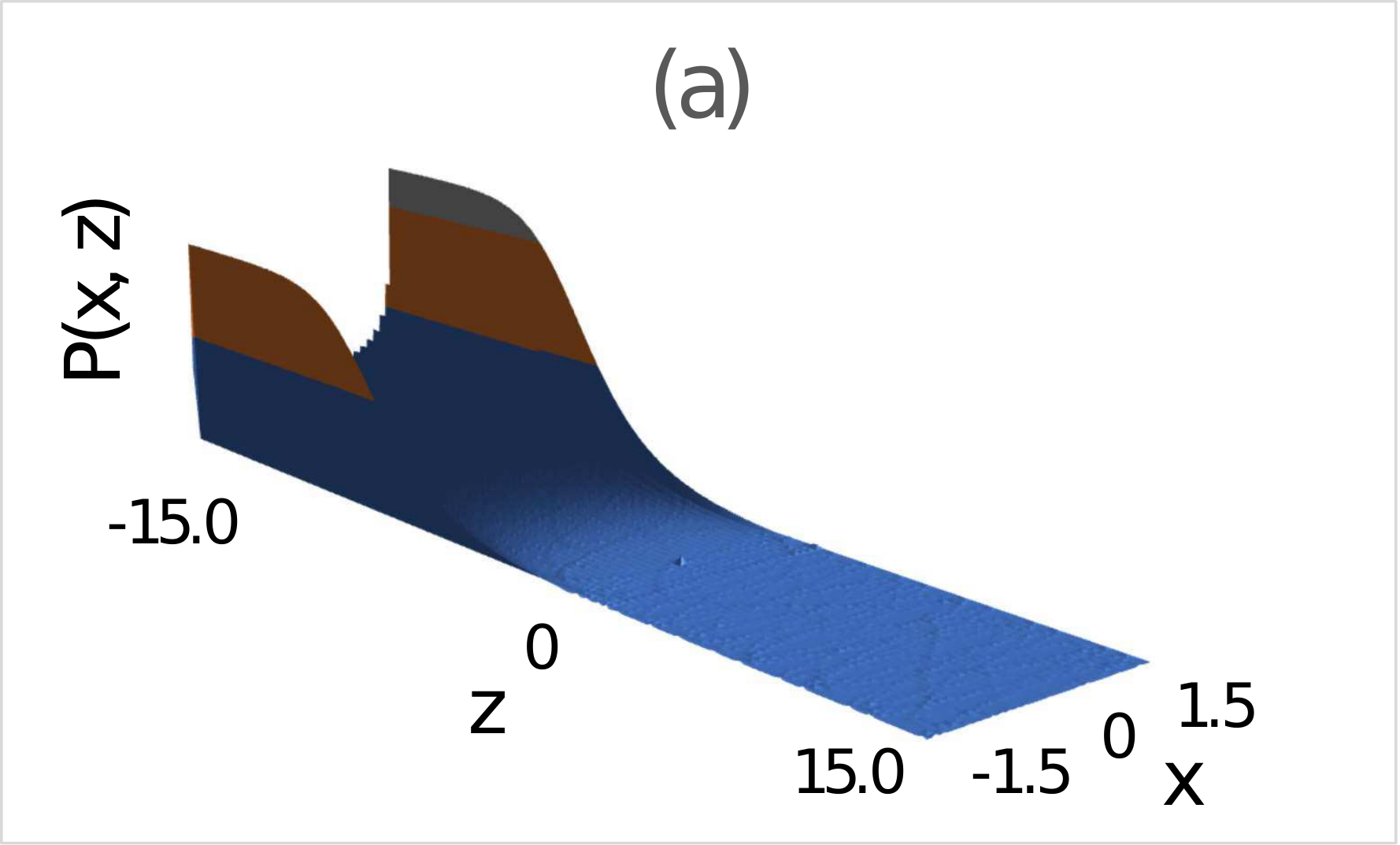} 
  \hfill
\includegraphics[width=0.5\textwidth]{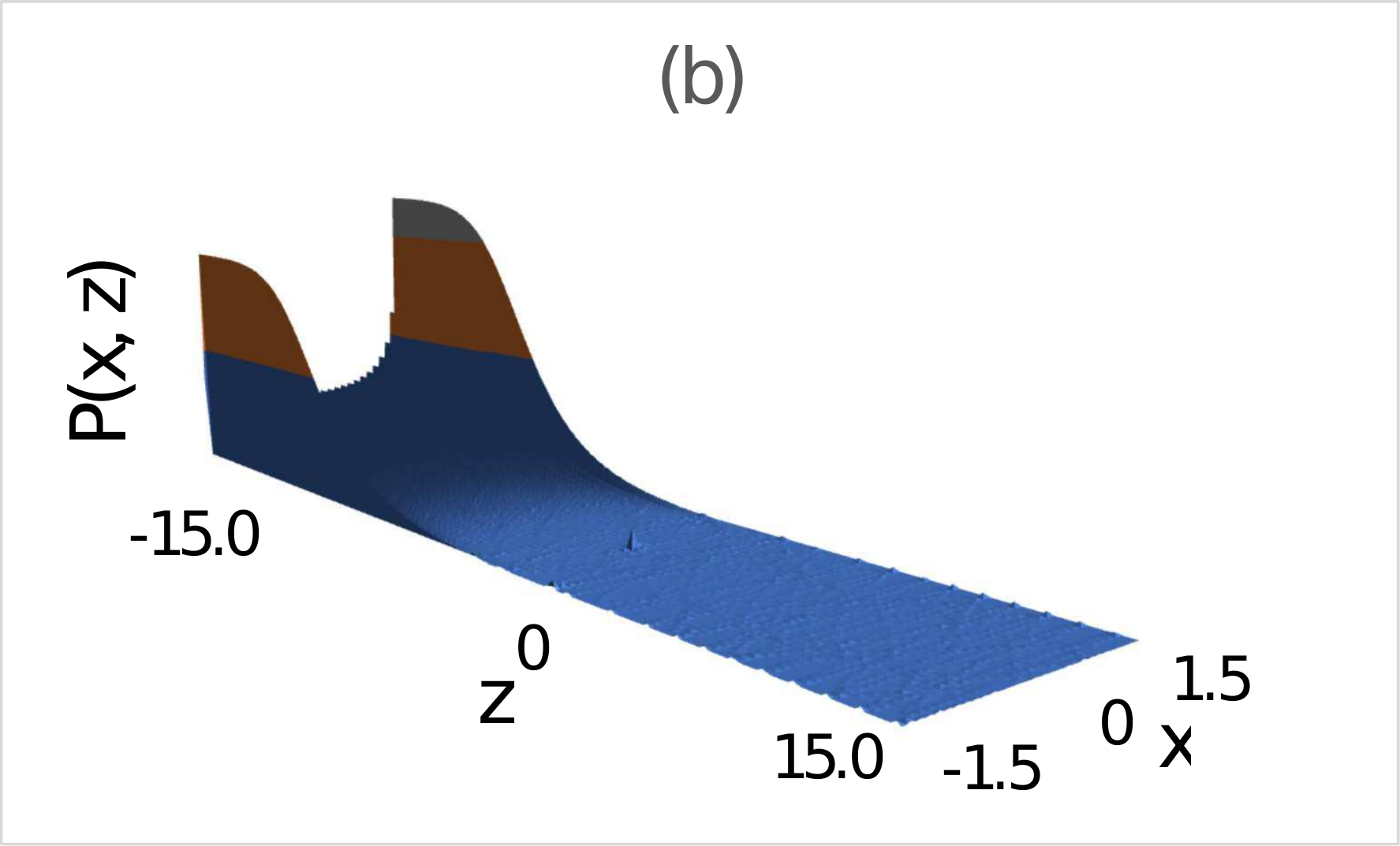} 
\hfill
\includegraphics[width=0.5\textwidth]{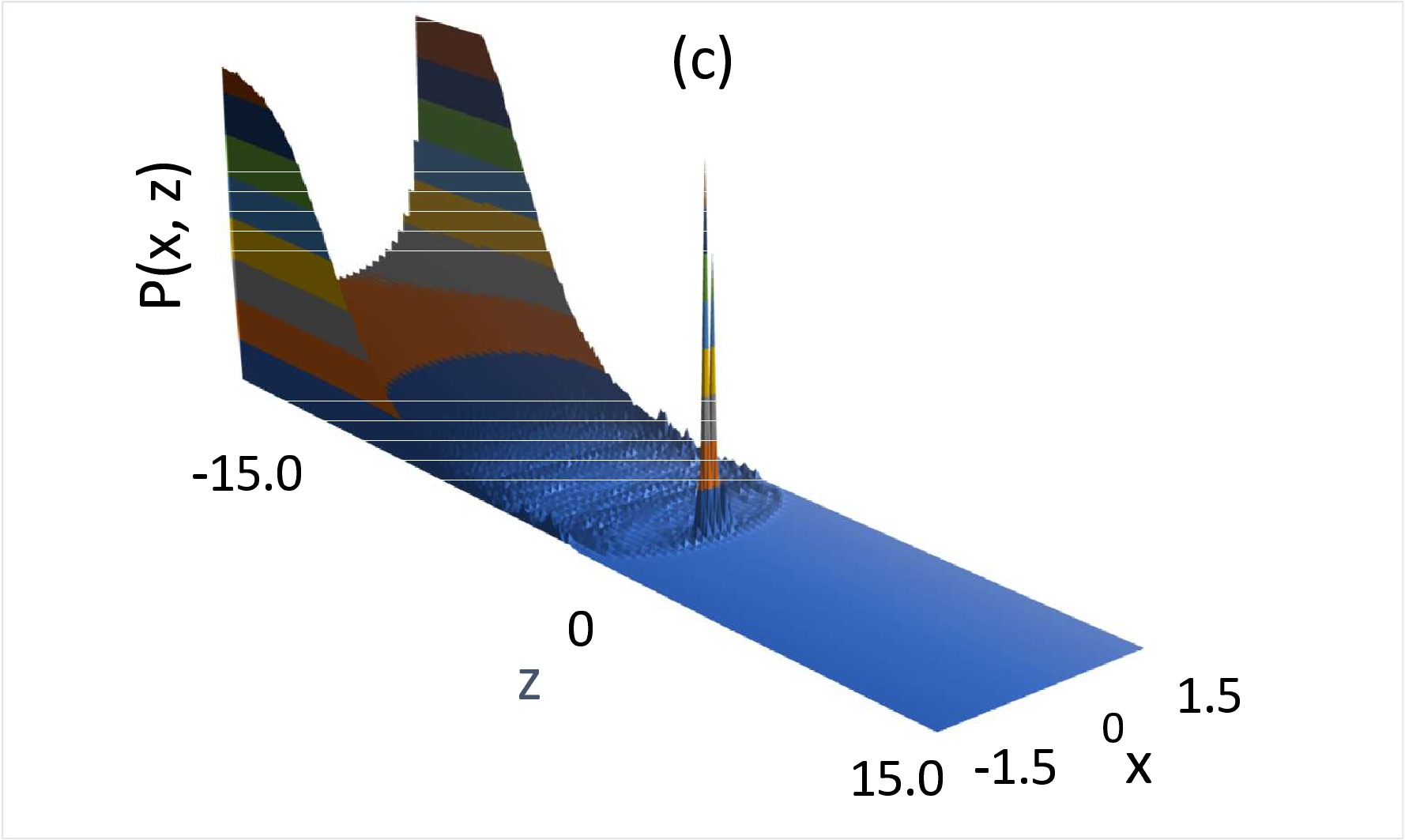} 
  \hfill
\includegraphics[width=0.5\textwidth]{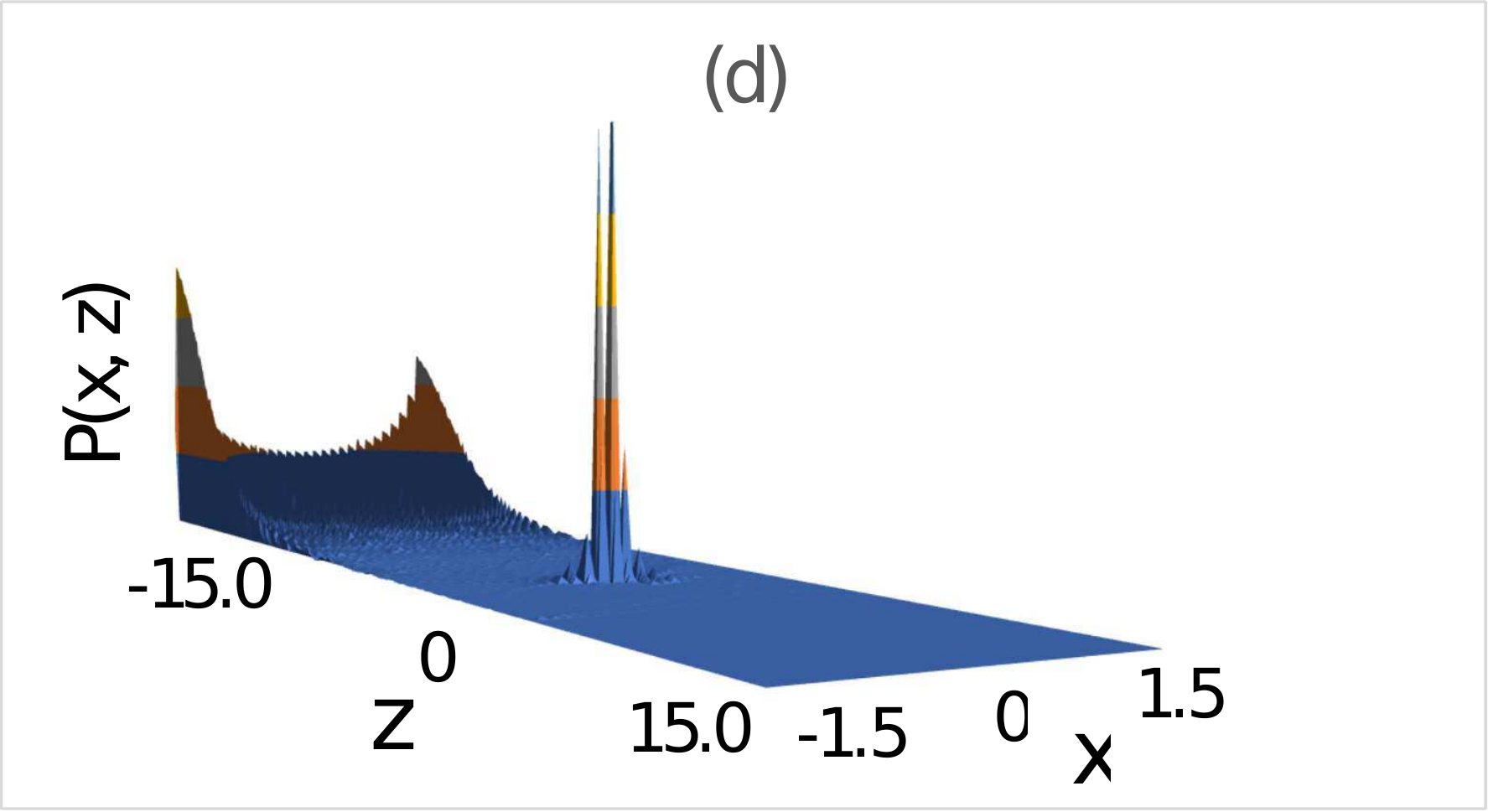} 
 \caption{\footnotesize (Color online) The relative probability density of the scattered particle as a function of its position $(x, z)$, for $y=0$, with $ E_{\perp}=1$, $E_{||}=1.0\times 10^{-5}$, and $L_z=0$ for the Yukawa potential for $V_0=0.01$ (a), $V_0=1.0$ (b), $V_0=9.0$ (c), and $V_0=100.0$ (d).  Here for simplicity we have considered the real case.}\label{Pro3D}
\end{figure}

\begin{figure}[!tbp] 
\centering
\includegraphics[width=0.5\textwidth]{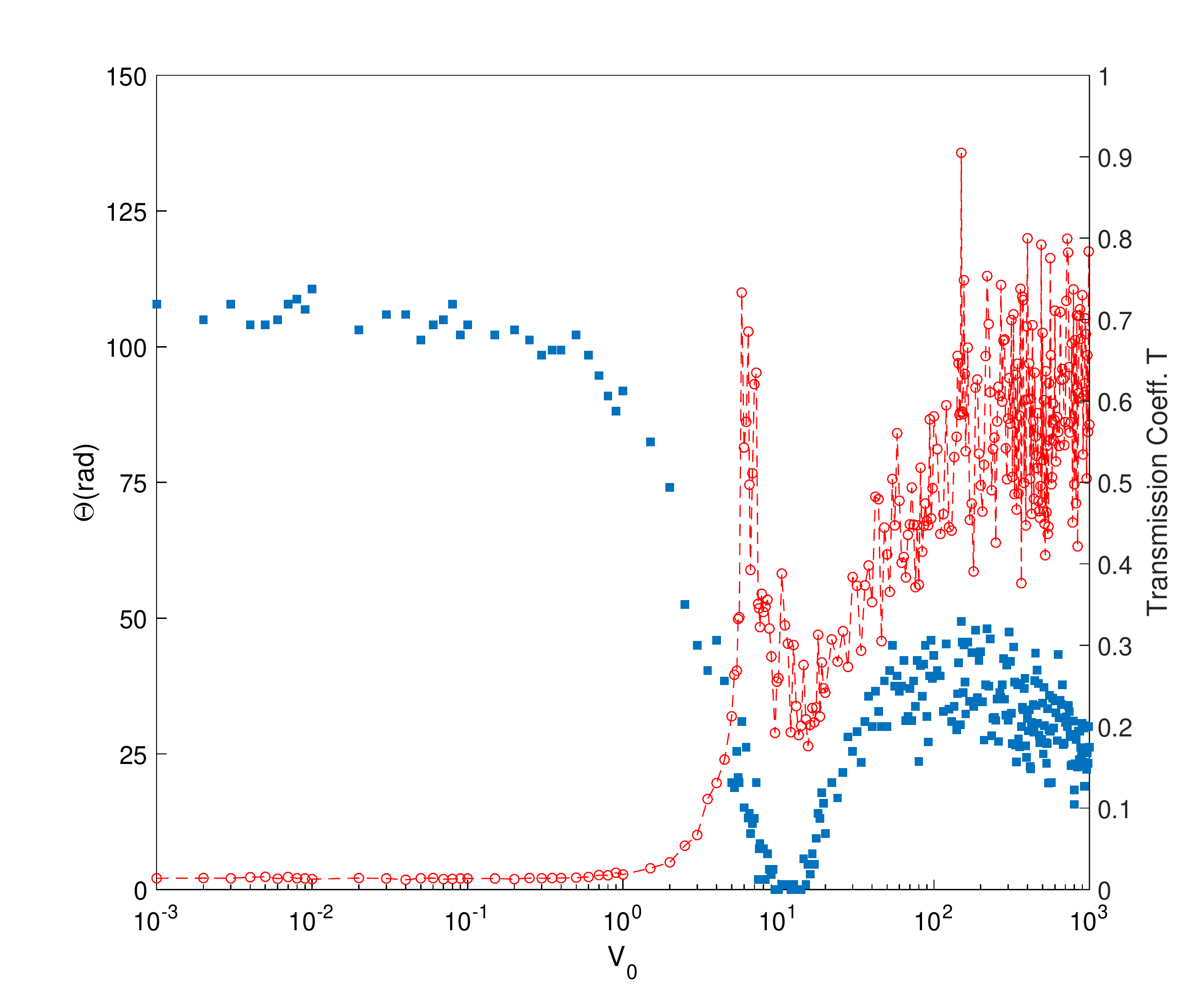} 
\hfill
\caption{ \footnotesize (Color online) A double plot of the scattering angle $\Theta$ (depicted on the left vertical axis and indicated in the graph as the (red) line with hollow circles) along with the transmission coefficient $T$ (represented on the right vertical axis and shown in the graph with (blue) filled squares) of a quasi-1D classical scattering, as functions of the interaction potential depth $V_0$ for the Yukawa potential.}\label{tetaV0waveguid}
\end{figure}

Now for the $\bold{classical}$ $\bold{ CIR}$. When the particle is constrained by the confining harmonic potential, it oscillates in the lateral ($x$ and/or $y$) direction(s) as it approaches the interaction potential. The result is a very complex movement with the scattering parameters $s$ and $v_0$ changing continuously in such a way that when the particle approaches the centre of the interaction potential, at times it finds itself able to get very close to the potential core, while at other times it finds itself constrained in a bound state corresponding to $V_{eff}(r)$. In all this, the particle has ample opportunity to revolve around the interaction potential centre. Eventually, however, it finds itself able to escape the effective interaction potential and depending on its escape angle, it would be scattered in the positive or negative $z$-directions while it continues to oscillate laterally. Confinement-induced resonance occurs when the conditions are most apt for the particle to be trapped within the interaction potential \cite{Note}.
To show this, we have plotted the relative probability density as a function of the particle's position $(x, z)$ for $y=0$ for several values of the interaction potential depth $V_0$ (Fig. \ref{Pro3D}) and also the variation of the scattering angle $\Theta$ in the $x-z$ plane with $V_0$ for $y=0$ (Fig. \ref{tetaV0waveguid}). For simplicity we have just considered the real case. 
In the complex case, the dimensions of space become large (6 dimensions for $x_r+ix_i, y_r+iy_i,$ and $z_r+iz_i $) and its geometry cannot be imagined. Although it is possible to define the variables of interest (for example, $\Theta$ in Fig. \ref{tetaV0waveguid}) for each 2D cut of the whole space giving 15 different planes, $(x_r,x_i), (x_r,z_r), ...$, this would make the whole picture quite complicated. The calculations for the real case, however, are deemed quite sufficient to demonstrate the physics of CIR in the classical case.

The relative probability density in Fig. \ref{Pro3D} was calculated by dividing the $x-z$ plane into small cells and recording the average number of times the particle passed through a particular cell on its trajectory.  As can be seen from the figure, at the turning points the probability density diverges in agreement with Fig. \ref{RelProCom}. On the other hand, for $V_0$'s near CIR or larger, $i.e.$, $V_0\geq V^{CIR}_0\simeq 9.0$ (since we have only considered the real case, the CIR position is different from what we have already seen in the complex case), we see peaks in the interaction region showing that, as in the quantum case, the incoming particle is trapped there, temporarily forming a pseudo-molecule. The figure demonstrates quite clearly that the particle is spending much longer times in the vicinity of the interaction potential compared to the case of direct collision; a picture that is expected of a Feshbach type resonance \cite{Moiseyev} and is consistent with the results obtained in the quantum case (see Fig. 8 of \cite{Saeidian08}). 

Fig. \ref{tetaV0waveguid} shows the rotation angle $\Theta$ along with the transmission coefficient $T$ plotted against $V_0$ for the Yukawa potential with $E_{\perp} = 1.0$ and $E_{||}= 10^{-5}$.   For small  $V_0$'s, $\Theta$ is very small implying that in most cases the particle does not rotate around the centre of the interaction potential. As $V_0$ approaches $V_0^{CIR}$, we observe a sharp peak in the value of $\Theta$. This clearly confirms our claim that when CIR occurs, the particle spends a great amount of time in the interaction region rotating about the centre of the interaction potential. Immediately after $V_0^{CIR}$, we observe a sharp decline in the value of $\Theta$. Now, as $V_0$ increases, $\Theta$ also increases slowly since a larger $V_0$ implies a stronger interaction potential forcing the particle to rotate more and more about the potential centre.

The above arguments indicate that the CIR that occurs in the classical case results from the spherical asymmetry of the combined potentials and is therefore of the Feshbach type.

\subsection{Scattering resonance in the classical case} \label{ScatResClassical}

\begin{figure}[tbp]
  \centering
  \includegraphics[width=0.5\textwidth]{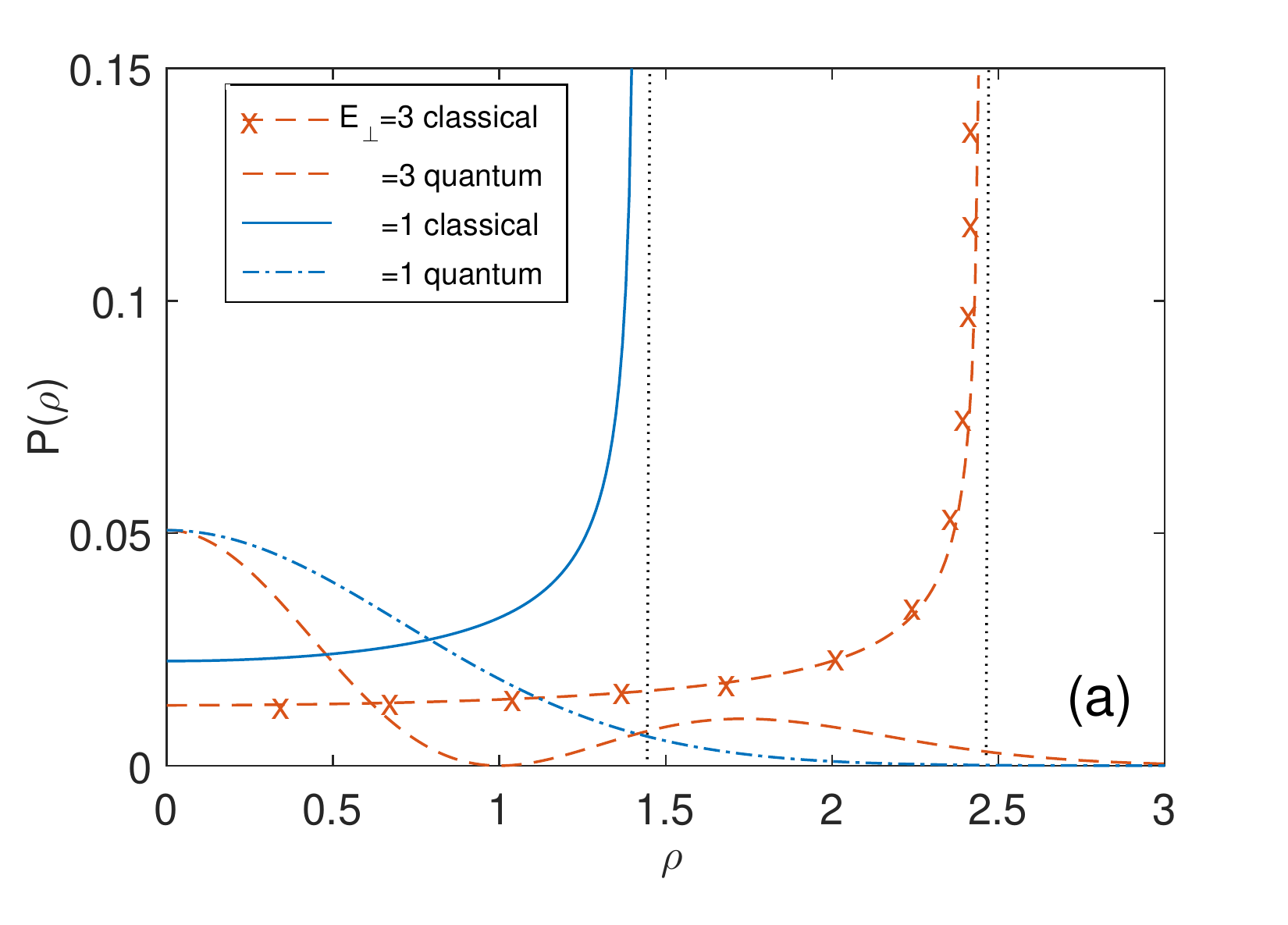}
  \hfill
  \includegraphics[width=0.5\textwidth]{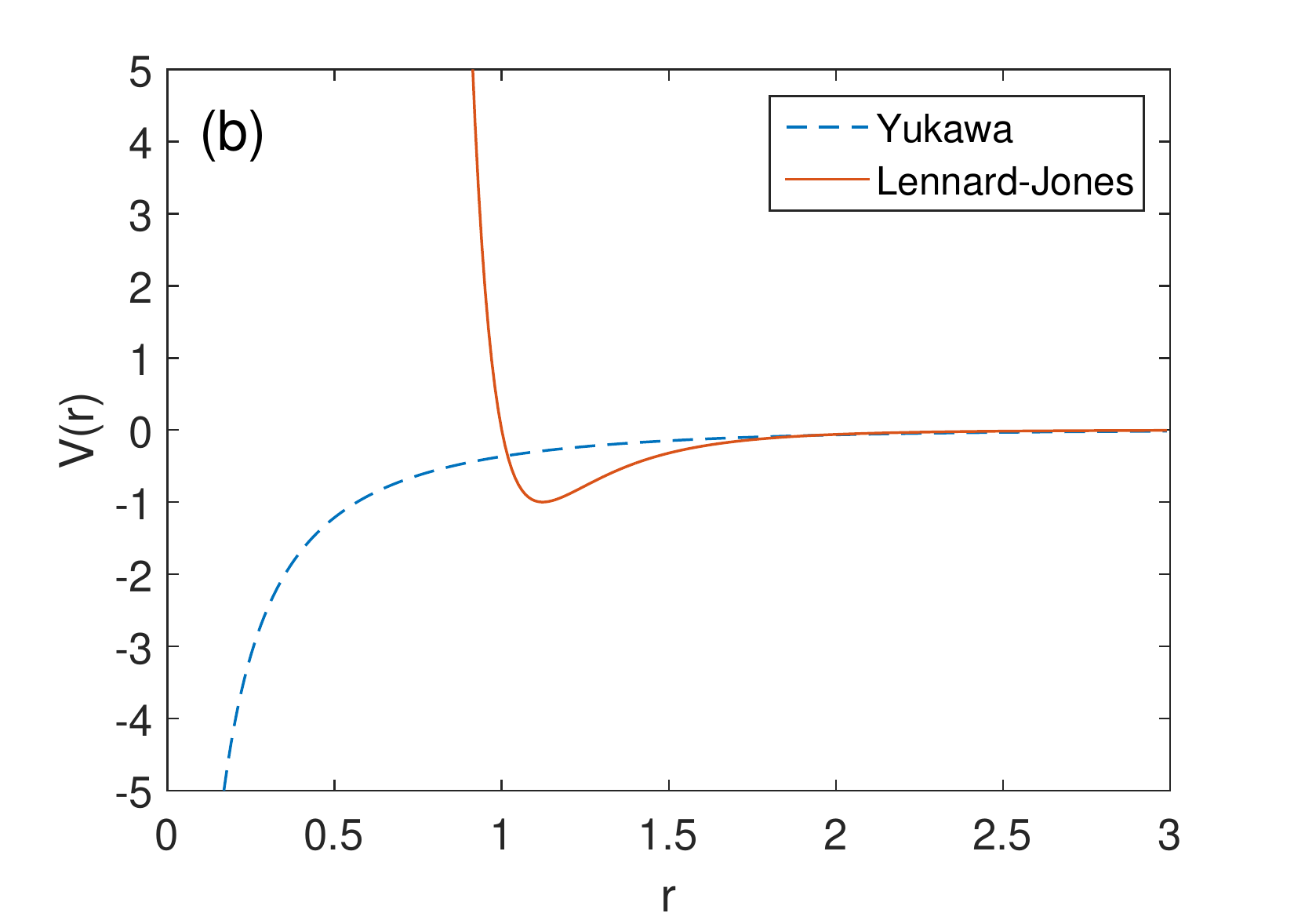} 
  \caption{ \footnotesize (Color online) (a) The unnormalized probability density function of a 2D harmonic potential for $E_{\perp}=1$ and $3$ in the classical and quantum cases. (b) A graph of the Yukawa and Lennard-Jones interaction potentials $V(r)$ versus the inter-particle distance $r$.}\label{ClassProbDist}
  \end{figure}

We start by explaining what happens to the ``position" of $T_{min}$ when the energy of the incoming particle ($E_{\perp}$ or $E_{||}$) is altered. Fig. \ref{ClassProbDist} (a) shows a plot of the particle probability density versus its distance from the longitudinal axis under 2D harmonic potential confinement in both the classical and quantum regimes. For the classical regime we have considered the real case ($i.e.$, $\rho_i=0$) for simplicity.  Examining the probability density shown in this figure it could be concluded that in the classical case, a larger impact parameter $s$ would be available at larger $E_{\perp}$s. On the other hand, according to Subsection VI.A., for a given incident particle energy, a larger $s$ requires a larger $V_0$ to trap the particle around the interaction potential explaining the shift to the right of $T_{min}$.  When a wider range of $s$ values are available to the incident particle, it can be trapped  around the interaction potential at a wider range of $V_0$ values. This effect may be seen as the flattening of the minimum in Fig. \ref{YukawaMulti5} (a). According to Table \ref{T1}, the shift to the right in $T_{min}$ occurs for both the Yukawa and Lennard-Jones potentials.

Now we are going to explain how the ``value"  of $T_{min}$ changes when the energy of the incoming particle ($E_{\perp}$ or $E_{||}$) is altered.

(i) According to Fig.  \ref{ClassProbDist}  (a), when $E_{\perp}$ increases, the particle moves, on average, away from the longitudinal axis spending most of its time near the border of the classically allowed region. Recalling the arguments put forward in Subsection VI. A., this would allow the particle to have access to a wider range of the impact parameter $s$. For each $s$ ($s*$, say), a $V_0$ may be found at which trapping around the interaction potential can occur. For that particular value of $V_0$, values of $s$ lower than $s*$ would allow the incident particle to penetrate the $s*$ potential bump. With a wider range of available $s$ values, the probability of penetration goes up, increasing the probability of the occurrence of resonance, and thus decreasing $T_{min}$. This is true for the Yukawa potential which is almost long-range, as shown in Fig.  \ref{ClassProbDist}  (b), which is a graph of both the Yukawa and Lennard-Jones interaction potentials. Therefore, when $E_{\perp}$  increases and moves the particle farther away from the longitudinal axis, it still feels the interaction potential. The Lennard-Jones potential, on the other hand, has a finite range so that when the increasing $E_{\perp}$ moves the particle away from the longitudinal axis, the particle will tend not to feel the L-J potential as much and this will result in an increase in the value of $T_{min}$.

(ii) In the classical case, as $E_{||}$ goes up, the interaction time is shortened and a larger value of $V_0$ is required for the probability of trapping to become significant. So, $V_0 (T_{min})$ shifts to the right. As for the change in the values of $T_{min}$, when $E_{||}$ is increased, the interaction time is reduced. On the other hand, since $E_{\perp}$ is fixed, the maximum amount of available $s$'s does not change. The combined effect is that the interaction potential has less chance to capture the particle and $T_{min}$ goes up. Figs. \ref{TVEparallel}(a) and \ref{TVEparallelLJ} (a) show that there is a maximum $E_{||}$ at which the $T_{min}$ dip is just discernible. The order of this maximum may be obtained through a simple calculation.  The approaching particle undergoes oscillations imposed by the 2D harmonic constraint. For CIR to occur, the distance the particle traverses along the longitudinal axis in one period of oscillation should at most be of the order of the range of the interaction potential. For $\omega = 1$, and an interaction potential with range $r_0 = 1$, we have $Max (E_{||})\sim 10^{-2}$. These calculations agree with the results shown in Figs. \ref{TVEparallel}(a) and \ref{TVEparallelLJ} (a).

We are now going to consider the effects of altering $L_z$ and $r_0$. As Fig. \ref{MultiLz} shows, changing $L_z$ does not seem to affect the position or the value of $T_{min}$. Keeping in mind that $E_{\perp}$ is fixed, changes in the rotations of the incoming particle do not seem to be important. This is because of the fact that in the classical case the 2D probability density (Fig. \ref{ClassProbDist} (a)) is not affected significantly by changing the $L_z$ values (at least for the values considered in this work).

As for the potential range, decreasing $r_0$ (Fig. \ref{MultiR0} (a)) results in a shift to the right and an increase in the value of $T_{min}$. For fixed $E_{||}$ and $E_{\perp}$, reducing $r_0$ will in effect reduce the interaction time. This will reduce the probability of capture, thus increasing the value of $T_{min}$. At the same time, these conditions would necessitate stronger interaction potentials $V_0$ for the capture of the particle around the interaction potential, hence, a $T_{min}$ shift to the right.

\subsection{Scattering resonance in the quantum case} \label{ScatResQuantum}

According to Figs. \ref{YukawaMulti5} and \ref{LJMulti}  and Table \ref{T1}, in the quantum case, as the transverse energy $E_{\perp}$ increases, the changes in $T_{min}$ are the same under either potential, $i.e.$, $T_{min}$ `increases' ($T_{min} \uparrow$) and its position shifts to the `left' ($T_{min} \leftarrow$) except for the Yukawa potential for which the shift is not discernible. Similarly, again according to Figs. \ref{TVEparallel} and \ref{TVEparallelLJ}, and Table \ref{T1}, as the longitudinal energy $E_{||}$ increases, the changes in $T_{min}$ are the same under either potential, $i.e.$, $T_{min}$ `increases' ($T_{min} \uparrow$) but there is no shift in its position ($T_{min} \times$).

\begin{figure}[h]
  \centering
  \includegraphics[width=0.5\textwidth]{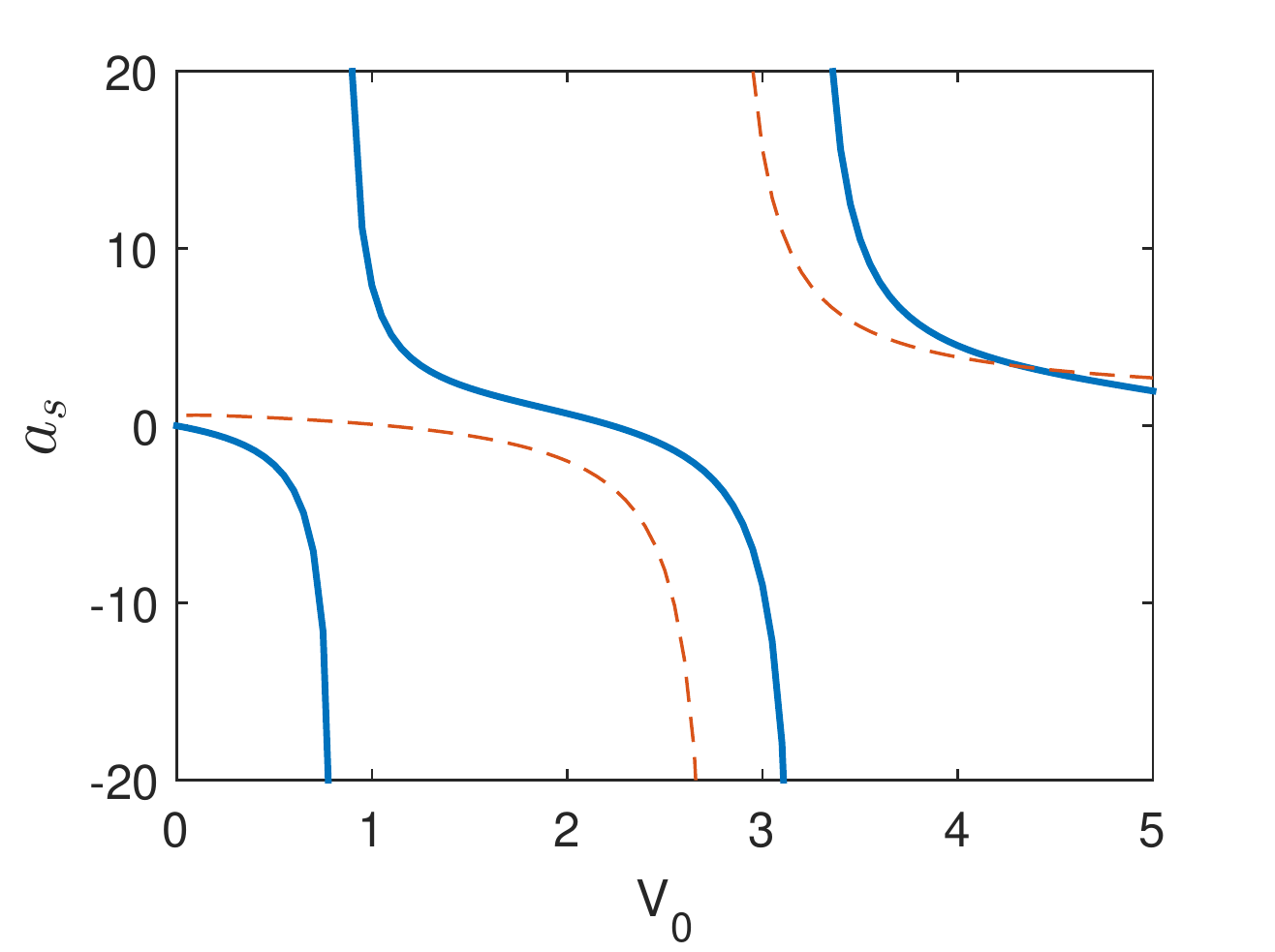} 
  \caption{ \footnotesize (Color online) The $s$-wave scattering length $a_s$ as a function of the potential depth $V_0$ for the Yukawa potential (Eq. \ref{yuk}) for $r_0=1$ (solid line) and the Lennard-Jones potential (Eq. \ref{LJp}) for $\sigma=1$  (dashed line).}\label{ScattLength}
  \end{figure}

As in the classical case, we start by explaining what happens to the position of $T_{min}$ when the energy of the incoming particle ($E_{\perp}$ or $E_{||}$) is altered. In the quantum case, the atom-atom interaction at low temperatures is described by the scattering lengths $a_l$ which are defined as $\lim_{k\rightarrow 0} \tan \delta_l/k^{2l+1}=-1/a_l^{2l+1}$.  Here $\delta_l$ is the phase shift of the $lth$ partial wave of the wave function in free space, due to the interaction potential, and $k$ is the wave number.  A larger scattering length corresponds to a stronger effective force by the interaction potential. Figure \ref{ScattLength} shows the scattering length $a_s$  as a function of the potential depth $V_0$ for the Yukawa and Lennard-Jones potentials. Each divergence in the graph corresponds to a resonance in free space. In our CIR calculations, we operate in an interval on the $V_0$  axis that is around one resonance ($V_0 = 0-2$ for Yukawa and $V_0 = 0-5$ for L-J). The observed behaviour is repeated periodically for other intervals. According to Fig. \ref{ClassProbDist} (a), in the quantum (and also the classical) case, when we increase $E_{\perp}$, the probability for the particle to be found near the axis decreases requiring a stronger effective force (corresponding to a larger scattering length) for CIR to occur. As can be seen in Fig. \ref{ScattLength}, for both Yukawa and Lennard-Jones interaction potentials, a larger scattering length is achieved at smaller $V_0$, $i.e.$, the position of $T_{min}$ shifts to the `left' ($T_{min} \leftarrow $ ) when we increase $E_{\perp}$.

For CIR to occur we just need to couple the eigenstates of the trap (which describe the transverse motion) with the interaction potential. We should keep changing $a_s$ until the energy of the bound state of the closed channel is near the threshold of the open channel. When $E_{\perp}$ is changed, the value of $a_s$ at which CIR occurs might also change. Fig. \ref{YukawaMulti5} (b) shows that $T_{min}$ occurs around $V_0 \sim 0.9$ and Fig. \ref{ScattLength} clearly shows that for the Yukawa potential, at $V_0 \sim 0.9$, even very large changes in $a_s$ can only result in a small change in $V_0$. Therefore, changes in $E_{\perp}$ would not shift the position of $V_0$ in a discernible manner ($T_{min} \times $). For the L-J potential, however, the changes are discernible (see Fig. \ref{TVEparallelLJ} (b)). 

When $E_{||}$ is increased, there is no shift in the position of $T_{min}$ ($T_{min} \times $). The reason for this is that for all the $E_{||}$ values chosen, the de Broglie wavelength for the particle motion along the longitudinal axis ($\lambda_{||} = 1/ \sqrt{2 E_{||}}$) is much larger than the range of the interaction potential. For example, for $E_{||} = 10^{-5}$ and $E_{||} = 10^{-2}$, these $\lambda$s are two and one orders of magnitude larger than the range of the potential (which is around one), respectively. Therefore, the interaction potential fails to differentiate between the various values of $\lambda$, resulting in the same $V_0(T_{min})$ for different $E_{||}$s.

Now we are going to explain how the value of $T_{min}$ changes when the energy of the incoming particle ($E_{\perp}$ or $E_{||}$) is altered.

(i) In the case of the transverse energy $E_{\perp}$, increasing $E_{\perp}$ corresponds to increasing the number of the open channels. The higher the number of the open channels, the higher the chances of the particle getting through, resulting in an increase in the transmission coefficient $T_{min}$.

(ii) When $E_{||}$ is increased, the interaction time is reduced and the interaction potential has less chance to couple the eigenstates of the confining potential. The result is less probability of capture. Therefore, $T_{min}$ goes up, as indicated in Figs. \ref{TVEparallel} (b) and \ref{TVEparallelLJ} (b).

We will now consider the effects of altering the range $r_0$ of the Yukawa potential. As Fig. \ref{MultiR0} (b) shows, for fixed $E_{||}$ and $E_{\perp}$, reducing the range of the interaction potential ($r_0$) will in effect reduce the interaction time. We would expect this to increase the value of $T_{min}$. On the other hand, in the quantum case, CIR occurs when there is a coupling between the ground state and the excited state of the confinement by the interaction potential. Since these two eigenstates are perpendicular to each other, the less the range of the interaction potential, the less it will be sensitive to their differences, and the more it will be able to couple them tending to decrease $T_{min}$. The latter dominates and $T_{min}$ goes down. As Fig. \ref{MultiR0}  (b) clearly indicates, if the range of the potential is decreased even further, the value of $T_{min}$ will approach zero (a one order of magnitude decrease in $r_0$ has resulted in four orders of magnitude decrease in $T_{min}$). This is in agreement with analytical results where Olshanii has demonstrated that $T_{min}$ becomes exactly equal to zero for the zero-range Huang potential \cite{Moore2003}. 

On the hand, when $r_0$ is reduced, the depth of the interaction potential needs to be increased for the bound state to even be formed (this can be easily seen in the case of a simple potential well). The result is $T_{min}$ shifting to the right.

\section{Summary and conclusion}

Using complex classical mechanics, we have carried out numerical studies of the classical analogue of the quantum confinement-induced resonance (CIR) for the scattering of two particles under a confining 2D harmonic potential; a so-called quasi-one-dimensional system. We used the Yukawa and Lennard-Jones potentials to emulate the interparticle interactions. In order to simulate the classical case in a manner that would lend itself to suitable comparison with the quantum case, we applied the  Bohr-Sommerfeld quantization rule to the confining potential to obtain the energy levels of the 2D harmonic potential and used these levels to represent the classical analogue of the quantum scattering channels.

We have observed what seems to be the qualitative equivalent of the quantum zero-energy CIR for the zero-energy limit of the longitudinal energy of the incident particle by observing a minimum in the transmission coefficient $T$. We investigated both the position $V_0 (T_{min})$ of the CIR ($V_0$ being the interaction potential depth) and the increase and decrease in the value of $T_{min}$ under different conditions of increasing / decreasing longitudinal and transverse energies of the incident particle for both the Yukawa and Lennard-Jones potentials, as summed up in Table \ref{T1}.

According to this table, in the quantum case, the value of $T_{min}$ increases, $i.e.$, there is more transition, in all cases of increasing $E_{||}$ or $E_{\perp}$. The table shows similar results in the classical case except for increasing $E_{\perp}$ under the Yukawa potential (the top row). We have put forward physical arguments for both similarities and differences. As for the position of $T_{min}$ of the $T-V_0$ graph, the table shows that in the quantum case there is no discernible shift in any cases of increasing $E_{\perp}$ or  $E_{||}$ except for increasing $E_{\perp}$ for L-J where there is a shift to the left, $i.e.$, $CIR$ occurs at a shallower interaction potential. In the classical case the shifts are all to the right ,$i.e.$, a stronger interaction potential is required for the resonance to occur. Here again we have endeavoured to provide sufficient physical arguments to explain the differences.

We extended our calculations to different values of the range $r_0$ of the Yukawa potential in both classical and quantum settings. In both cases $T_{min}$ shifts to the right as $r_0$ decreases.  However, in the quantum case the value of $T_{min}$ decreases while in the classical case it increases.  

We also carried out classical calculations for different values of $L_z$ while keeping $E_{\perp}$ and  $E_{||}$ fixed and found that changing  $L_z$ made no significant difference in the results. To the best of our knowledge, no quantum calculations have been made so far for $L_z \ne 0$ to compare our classical results with except for the zero-range Huang potential where Olshanii has demonstrated that there is no CIR for $L_z \ne 0$.

The analogy seems to be complete. We hope that this work will provide a better and deeper understanding of the physics of scattering under 2D confinement and the resulting resonance in the quantum case.   Our future work will include numerical calculations of $L_z \ne 0$ in the quantum case and their comparison with classical calculations.



\begin{thebibliography}{1}
\bibitem{Note} If the particle is not trapped, it will certainly be transmitted, $T = 1$. However, if it is trapped with probability $p$, the probability of transmission will be $T = 1- p . p_r$ where $p_t$ and $p_r = 1 - p_t$ are the probability of transmission and reflection after trapping, respectively. Therefore, the more the probability of the particle being trapped, the higher will be the probability of it being reflected. CIR corresponds to the maximum probability of trapping manifesting itself in $T- V_0$ graphs as $T_{min}$.
\end{thebibliography}
\end{document}